\documentclass[format=acmsmall, review=false, screen=true]{acmart}

\usepackage{booktabs} 
\usepackage{multirow}
\usepackage{subcaption}
\usepackage{amsmath}
\usepackage{amsfonts}
\usepackage{amssymb}
\usepackage{enumitem}

\usepackage{rotating, graphicx}
\usepackage{array}
\usepackage{gensymb} 
\usepackage[ruled]{algorithm2e} 
\usepackage{textcomp}

\SetAlFnt{\small}
\SetAlCapFnt{\small}
\SetAlCapNameFnt{\small}
\SetAlCapHSkip{0pt}
\IncMargin{-\parindent}
\setlist[enumerate]{leftmargin=*}

\renewcommand\footnotetextcopyrightpermission[1]{} 
\acmVolume{0}
\acmNumber{0}
\acmArticle{0}
\acmYear{0}
\acmMonth{0}
\copyrightyear{2017}

\setcopyright{acmlicensed}

\acmDOI{0000001.0000001}

\received{June 2018}
\received[revised]{2018}

\newcommand\blfootnote[1]{%
	\begingroup
	\renewcommand\thefootnote{}\footnote{#1}%
	\addtocounter{footnote}{-1}%
	\endgroup
}

\begin{document}
\title[Security and Privacy Approaches in MR]{Security and Privacy Approaches in Mixed Reality:\newline A Literature Survey}  

\author{Jaybie A. de Guzman}
\affiliation{%
  \institution{University of New South Wales}
  \city{Sydney}
  \state{NSW}
  \postcode{2052}
  \country{Australia}
  }
  \affiliation{%
  \institution{Data 61, CSIRO}
 \streetaddress{13 Garden St}
 \city{Eveleigh} 
 \state{NSW}
   \postcode{2015}
 \country{Australia}
  }
\email{j.deguzman@student.unsw.edu.au}
\author{Kanchana Thilakarathna} 
 \affiliation{%
  \institution{University of Sydney}
  }
 \affiliation{%
 \institution{Data 61, CSIRO}
 }
\author{Aruna Seneviratne}
\affiliation{%
  \institution{University of New South Wales}
  }
    \affiliation{%
  \institution{Data 61, CSIRO}
  }

\begin{abstract}
Mixed reality (MR) technology development is now gaining momentum due to advances in computer vision, sensor fusion, and realistic display technologies. With most of the research and development focused on delivering the promise of MR, there is only barely a few working on the privacy and security implications of this technology. This survey paper aims to put in to light these risks, and to look into the latest security and privacy work on MR. Specifically, we list and review the different protection approaches that have been proposed to ensure user and data security and privacy in MR. We extend the scope to include work on related technologies such as augmented reality (AR), virtual reality (VR), and human-computer interaction (HCI) as crucial components, if not the origins, of MR, as well as numerous related work from the larger area of mobile devices, wearables, and Internet-of-Things (IoT). We highlight the lack of investigation, implementation, and evaluation of data protection approaches in MR. Further challenges and directions on MR security and privacy are also discussed.
\end{abstract}

\ccsdesc[500]{Human-centered computing~Mixed / augmented reality}
\ccsdesc[300]{Security and privacy~Privacy protections}
\ccsdesc[300]{Security and privacy~Usability in security and privacy}
\ccsdesc[100]{General and reference~Surveys and overviews}

\keywords{Mixed Reality, Augmented Reality, Privacy, Security}

\thanks{This work is supported by ...

  Authors' addresses: ...}

\maketitle

\renewcommand{\shortauthors}{J. de Guzman et al.}

\section{Introduction}\label{sec:intro}

\blfootnote{A published verion of this work is available at \url{https://doi.org/10.1145/3359626}. Please cite the published version as: \newline
Jaybie A. De Guzman, Kanchana Thilakarathna, and Aruna Seneviratne. 2019. Security and Privacy Approaches in Mixed Reality: A Literature Survey. ACM Comput. Surv. 52, 6, Article 110 (October 2019), 37 pages.}

\begin{figure}[t]
	\centering
	\includegraphics[width=0.85\textwidth]{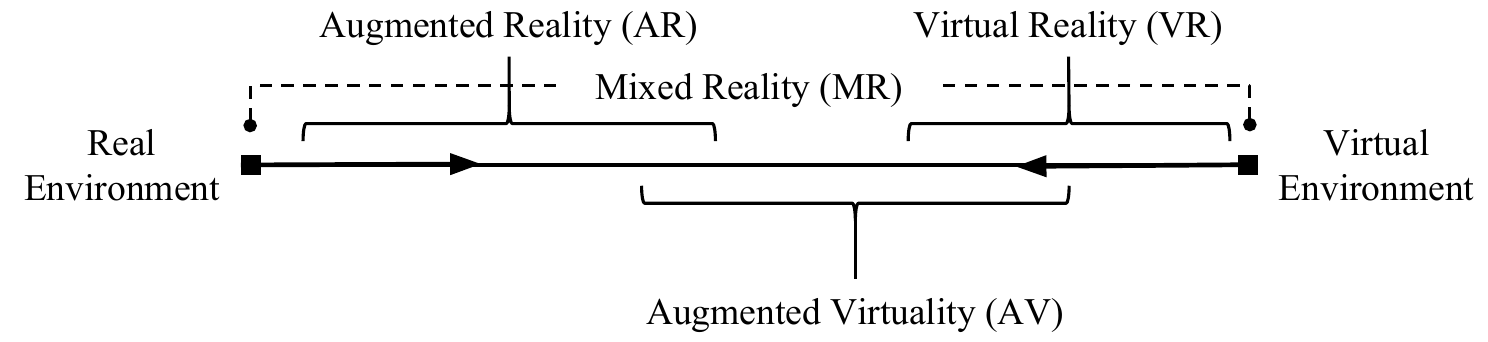}
	\vspace{-2mm}
	\caption{Milgram and Kishino's Reality and Virtuality Continuum \cite{milgram1994taxonomy}, and \textit{relative} positions of where AR, VR, and AV are along the continuum. AR sits near the \textit{real environment} as its primary intention is to \textit{augment} synthetic objects to the physical world while VR sits near the \textit{virtual environment} with varying degree of real-world information being feed in to the virtual experience. A third intermediary type, \textit{augmented virtuality}, sits in the middle where actual real-world objects are integrated into the virtual environment and thus intersecting with both AR and VR.}
	\label{fig:Milgram}
	\vspace{-2.5mm}
\end{figure}

\textit{Mixed reality} (MR) was used to pertain to the various devices -- specifically, displays -- that encompass the \textit{reality-virtuality continuum} as seen in Figure \ref{fig:Milgram} \cite{milgram1994augmented}. This means that \textit{augmented reality} (AR) systems and \textit{virtual reality} (VR) systems are MR systems but, if categorized, will lie on different points along the continuum. Presently, mixed reality has a hybrid definition that combines aspects of AR and VR to deliver rich services and immersive experiences \cite{mixedreality}, and allow interaction of real objects with synthetic virtual objects and vice versa. By combining the synthetic presence offered by VR and the extension of the real world by AR, MR enables a virtually endless suite of applications that is not offered by current AR and VR platforms, devices, and applications.

Advances in computer vision -- particularly in object sensing, tracking, and gesture identification -- , sensor fusion, and artificial intelligence has furthered the human-computer interaction as well as the machine understanding of the real-world. At the same time, advances in 3D rendering, optics -- such as projections, and holograms --, and display technologies have made possible the delivery of realistic virtual experiences. All these technologies make MR possible. As a result, MR can now allow us to interact with machines and each other in a totally different manner: for example, using gestures in the air instead of swiping in screens or tapping on keys. The output of our interactions, also, will no longer be confined within a screen. Instead, outputs will now be mixed with our real-world experience, and possibly sooner we may not be able to tell which is real and synthetic. Recently released MR devices such as Microsoft's Hololens and the Magic Leap demonstrates what these MR devices can do. They allows users to interact with holographic augmentations in a more seamless and direct manner. 

Most of the work on MR for the past two decades have been focused on delivering the necessary technology to make MR a possibility. As the necessary technology is starting to mature, MR devices, like any other technology, will become more available and affordable. Consequently, the proliferation of these devices may entail security and privacy implications which may not yet be known. For example, it has been demonstrated how facial images captured by a web camera can be cross-matched with publicly available online social network (OSN) profile photos to match the names with the faces and further determine social security numbers \cite{acquisti2011}. 
With most MR devices coming out in a wearable, i.e. head-mounted, form-factor and having at least one camera to capture the environment, it will be easy to perform such facial matching tasks in the wild without the subjects knowing it. Security and privacy, most often than not, comes as an afterthought, and we are observing a similar trend in MR as further shown in this survey paper.

\begin{figure}[t]
	\centering
	\includegraphics[width=\textwidth]{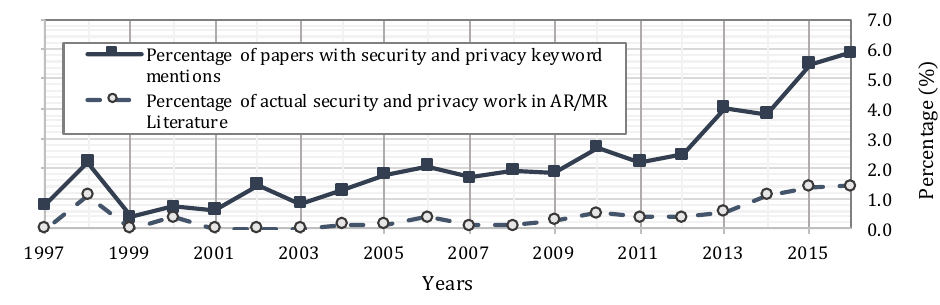}
	\vspace{-6mm}
	\caption{Statistics from Scopus: 
		The overall percentage of papers with ``security'' and ``privacy'' keywords, and, after removing papers with just keyword mentions, the percentage of actual papers on security and privacy. }
	\label{fig:scopus_statistics}
	\vspace{-2.5mm}
\end{figure}

To systematically capture the trend, we used the search tool of Scopus to initially gather AR and MR literature. We further identified works with security and privacy from this gathered list. 
Although, Scopus does not index literature from all resources, particularly in security and privacy research, the search tool can include \textit{secondary documents} that have been cited by Scopus-indexed documents which now effectively includes most security and privacy literature such as those from  USENIX Security Symposium, the Internet Society's Networks and Distributed System Security (NDSS) Symposium, and so on. 
Figure \ref{fig:scopus_statistics} shows the yearly percentage of these papers. Despite the increasing percentage from 0.7\% in 1997 to 5.8\% in 2016, most only mention security and privacy, and only a few (1.42\% for 2016) are actually discussing the \textit{impacts}, or presenting security and privacy approaches \textit{applied to}, or \textit{using} AR/VR/MR. 
Nonetheless, we supplement the gathered works from Scopus by separately searching for AR/VR/MR works with security and privacy from Google Scholar, IEEE Xplore, ACM Digital Library, and other specific venues covering computer vision, human-machine interfaces, and other related technologies.

\subsection*{Previous Surveys and Meta-analyses}

Early surveys on AR and MR, have been focused on categorizing the existing technologies then. In 1994, a taxonomy for classifying mixed reality displays based on the user-interface -- from monitor-based video displays to completely immersive environments -- were presented and these devices were plotted along a reality-virtuality continuum \cite{milgram1994taxonomy}. On the other hand, in contrast to this one-dimensional continuum, two different classifications for mixed reality were also presented: (1) a two-dimensional categorization of \textit{shared space} or collaborative mixed reality technologies according to concepts of \textit{transportation}\footnote{Transportation refers to the extent to which the users are transported from their physical locality to a virtual or remote space.} and \textit{artificiality}\footnote{Artificiality is the extent to which the user space has been synthesized from a real physical space.}, and (2) a one-dimensional classification based on \textit{spatiality}\footnote{Spatiality is the extent to which properties of natural space and movement is supported by the shared space.} \cite{benford1996shared}.

Succeeding endeavours have focused on collecting all relevant technologies necessary to AR and VR. The various early challenges -- such as matching the real and virtual displays, aligning the virtual objects with the real world, and the various errors that needs to be addressed such as optical distortion, misalignment, and tracking -- have been discussed in broad \cite{azuma1997survey}. It was complemented with a following survey that focuses on the enabling technologies, interfacing, and visualization \cite{azuma2001recent}. A much more recent survey updated the existing challenges to the following: performance, alignment, interaction, mobility, and visualization \cite{rabbi2013survey}. Another one looked into a specific type of AR, mobile AR, and looked into the different technologies that enable mobility with AR \cite{chatzopoulos2017mobile}. Lastly, a review of the various head-mounted display (HMD) technologies for consumer electronics \cite{kress2013HMD} was also undertaken. While all of these different challenges and technologies are important to enable AR, none of these survey or review papers have focused onto the fundamental issues of security and privacy in AR or MR.

A few others have pointed out the non-technical issues such as ethical considerations \cite{heimo2014ethical} and value-sensitive design approaches \cite{friedman2000value} that pushes to consider data ownership, privacy, secrecy, and integrity. A much recent work emphasized the three aspects for protection in AR -- \textit{input}, \textit{data access}, and \textit{output} -- over varying system complexity (from single to multiple applications, and, eventually, to multiple systems) \cite{Roesner2014}. In supplement, we expand from these three aspects, and include \textit{interaction} and \textit{device protection} as equally important aspects and dedicate separate discussions for both. A very recent survey collected offensive and defensive approaches on wearables \cite{shrestha2017offensive} and it included some strategies for wearable eye-wears and HMDs. In this work, we expand the coverage to include all MR platforms and setups (including non-wearables) and present a more specific and, yet, wider exposition of MR security and privacy approaches.

\subsection*{Contributions}
To the best of our knowledge, this is the first survey on the relevant security and privacy challenges and approaches on mixed reality. To this end, this work makes the following contributions:
\begin{enumerate}
\item We provide a \textit{data-centric} categorization of the various works which categorizes them to five major aspects, and, further, to subcategories, and present generic system block diagrams to capture the different mechanisms of protection.
\item We include a collection of other relevant work not necessarily directed to mixed reality but is expected to be related to or part of the security and privacy considerations for MR.
\item Lastly, we identified the target security and privacy properties of these approaches and present them in a summary table to show the distribution of strategies of protection among the properties.
\end{enumerate}

Before proceeding to the review of the various security and privacy work, we clarify that 
we do not focus on network security and related topics. We will rely on the effectiveness of existing security and privacy measures that protects the communication networks, and data transmission, in general.

The rest of the survey proceeds as follows. 
\S\ref{sec:properties&categorization} lists the different security and privacy properties, and explains the categorization used to sort the different approaches. The details of these approaches and which properties they address are discussed in \S\ref{sec:s&p}. 
Lastly, in \S\ref{sec:challenges}, the open challenges and future directions are discussed before finally concluding this survey in \S\ref{sec:conclusion}.

\section{Defining and Categorizing Security and Privacy in Mixed Reality}\label{sec:properties&categorization}
In this section, we first present the important security and privacy requirements that needs to be considered in designing or using MR. Then, we present an an overall categorization that focuses on the \textit{flow} of data within an MR system.

\subsection{General Security and Privacy Requirements}
We derive security and privacy properties from three models and combine them to have an over-arching model from which we can refer or qualify the different approaches (both defensive and offensive strategies) that will be discussed in this survey paper. Table \ref{tab:unified-model} lists the thirteen combined security and privacy properties from Microsoft's \textit{Security Development Lifecycle} \cite{howard2006security}, PriS \cite{kalloniatis2008addressing}, and LINDDUN \cite{deng2011privacy} and their corresponding threats. The first six are security properties while the remaining are considered as privacy properties. The confidentiality property is the only common among the three models and is considered as both a security and privacy property. Obviously, since the Microsoft's SDL focuses primarily on security, all of its associated properties target security. PriS has a roughly balanced privacy and security targeted properties. On the other hand, LINDUNN's properties are privacy-targeted, and are further categorized into hard privacy (from confidentiality down to plausible deniability) and soft privacy (the bottom two properties).

\begin{table}
	\centering
	\scriptsize
	\caption{Combined general security and privacy properties and their corresponding threats}
	\label{tab:unified-model}
	\setlength{\extrarowheight}{.5em}
	\begin{tabular}{@{}llp{1.3cm}p{1.3cm}p{1.3cm}c@{}}
		\toprule
		\multirow{2}{*}{\textbf{Property}} & \multirow{2}{*}{\textbf{Threat}} & \multicolumn{3}{c}{\textbf{Model}} \\ \cmidrule{3-5}
		&& Microsoft's SDL 2006 & PriS 2008 & LINDDUN 2011 & \\ \midrule
			Integrity & Tampering  &\centering \checkmark& & &\\
		Non-repudiation & Repudiation  & \centering \checkmark& & &\\
		Availability & Denial of Service  & \centering \checkmark& &&\\ 
		Authorization & Elevation of Privilege  & \centering \checkmark& \centering \checkmark& &\\
		Authentication & Spoofing  &\centering \checkmark& \centering \checkmark&&\\ 
		Identification & Anonymity & & \centering \checkmark &&\\
		Confidentiality & Disclosure of Information  &\centering \checkmark&\centering \checkmark & \centering \checkmark&\\ 
		Anonymity \& Pseudonymity & Identifiability   & &\centering \checkmark & \centering \checkmark&\\
		Unlinkability & Linkability  & & \centering \checkmark & \centering \checkmark &\\
		Unobservability \& Undetectability & Detectability && \centering \checkmark&\centering \checkmark &\\
		Plausible Deniability& Non-repudiation && &\centering \checkmark &\\
		Content Awareness & Unawareness & & & \centering \checkmark&\\
		Policy \& Consent Compliance& Non-compliance  && &\centering \checkmark & \\
		\bottomrule
	\end{tabular}
	\vspace{-3mm}
\end{table}

It is interesting to note that some security properties are conversely considered as privacy threats, such as non-repudiation and plausible deniability. This highlights the differences in priority that an organization, user, or stakeholder can put into these properties or requirements. However, these properties are not necessarily mutually exclusive and can be desired at the same time. Moreover, the target entity (or element) adds another dimension to this list. Specifically, these properties can be applied to the following entities or data flow elements: pieces of data, data flow, process, and data storage. We highlight these co-existence as well as example target entities as we go through each of these properties one-by-one.

\begin{enumerate}
	\item \textit{Integrity} -- The data, flow, or process in MR is not and cannot be tampered or modified. This is to ensure that, for example, the visual targets are detected correctly and the appropriate augmentations are displayed accordingly. No unauthorized parties should be able to modify any element in MR.
	\item \textit{Non-repudiation} -- Any modification, or generation of data, flow, or process cannot be denied especially if the entity is an essential one or an adversary was able to perform such modifications or actions. When necessary, the modifier or generator of the action should be identified and cannot deny that it was their action. In privacy, however, the converse is desired.
	\item \textit{Availability} -- All necessary data, flow, or process for MR should be available in order to satisfy and accomplish the targeted or intended service. An adversary should not be able to impede the availability of these entities or resources.
	\item \textit{Authorization and Access Control} -- All actions should be originated from authorized and verifiable parties. The same actions are also actuated according to their appropriate access privileges. For example, a navigation application requires GPS information but does not necessarily require camera view. Or only the augmentations from applications that have been authorized to deliver augmentations should be rendered.
	\item \textit{Identification} -- All actions should be identified to the corresponding actor, i.e. user or party. In a security context, this is interrelated with authorization and authentication properties. Verified identities are used for authorizing access control. Unidentified parties can be treated as adversaries to prevent unidentifiable and untraceable attacks. In sensitive situations, e.g. an attack has occurred, anonymity is not desired.
	\item \textit{Authentication} -- Only the legitimate users of the device or service should be allowed to access the device or service. Their authenticity should be verified through a feasible authentication method. Then, identification or authorization can follow after a successful authentication.
	\item \textit{Confidentiality} -- All actions involving sensitive or personal data, flow, or process should follow the necessary authorization, and access control policies. Parties that are not authorized should not have access to these confidential entities. All entities can be assumed as confidential especially personal and re-identifiable data.
	\item \textit{Anonymity \& Pseudoymity} -- Users should be able to remove their association or relationship to the data, flow, or process. Likewise, a pseudonym can be used to link the entities but should not be linked back to user identities. Moreover, an adversary should not be able to identify the user from combinations of these entities. However, in the security context, anonymity is not desired especially when adversaries need to be identified.
	\item \textit{Unlinkability} -- Any link or relationship of the user or party to the data, flow, or process as well as among the entities (e.g. data to data, data to flow, and so on) cannot be distinguished by an adversary.
	\item \textit{Undetectability \& Unobservability} -- Any entities' existence cannot be ensured or distinguished by an attacker. Or an entity can be deemed unobservable or undetectable by an adversary that does not require it. Or the entity cannot be distinguished from randomly generated entities. For example, an MR game like PokemonGo needs access to the camera view of the user device and determine the ground plane to place the Pokemon on the ground as viewed by the user, but the game does not need to know what the other objects within view are.
	\subitem These two properties can be extended to include \textit{latent privacy} protection. It is the protection of entities that are not necessitated by an application or service but can be in the same domain as their target entities. This includes \textit{bystander privacy}.
	\item \textit{Plausible Deniability} -- The user or party should be able to deny their relationship to the entities, or of among themselves. This is a converse of the non-repudiation security property which is actually the corresponding threat for plausible deniability (or repudiation). However, this property is essential when the relationship to personal sensitive data should not be compromised, while non-repudiation is essential if, for example, an adversarial action should not be denied by an attacker.
	\item \textit{Content Awareness} -- The user should be aware of all data, flows or processes divulged especially those of sensitive entities. And that they should be aware that they have released the necessary amount of information or if it was too much.
	\item \textit{Policy and consent Compliance} -- The system should follow or policies specified that aims to protect the user's privacy or security. There should be a guarantee that systems, especially third-party applications or services, follow these policies and considers the consent allowed to them.
\end{enumerate}

For every approach that will be discussed in the next section (\S\ref{sec:s&p}), we will identify which properties they are trying to address. Moreover, there are other `soft' properties (i.e. \textit{reliability}, and \textit{safety}.) that we will liberally use in the discussions. We categorize these approaches in a \textit{data-} and \textit{data flow}-centric fashion which is explained in the next subsection.

\subsection{Categorizing the Threats}\label{ssec:categorization}

\begin{figure*}[t]
\centering
\includegraphics[width=\textwidth]{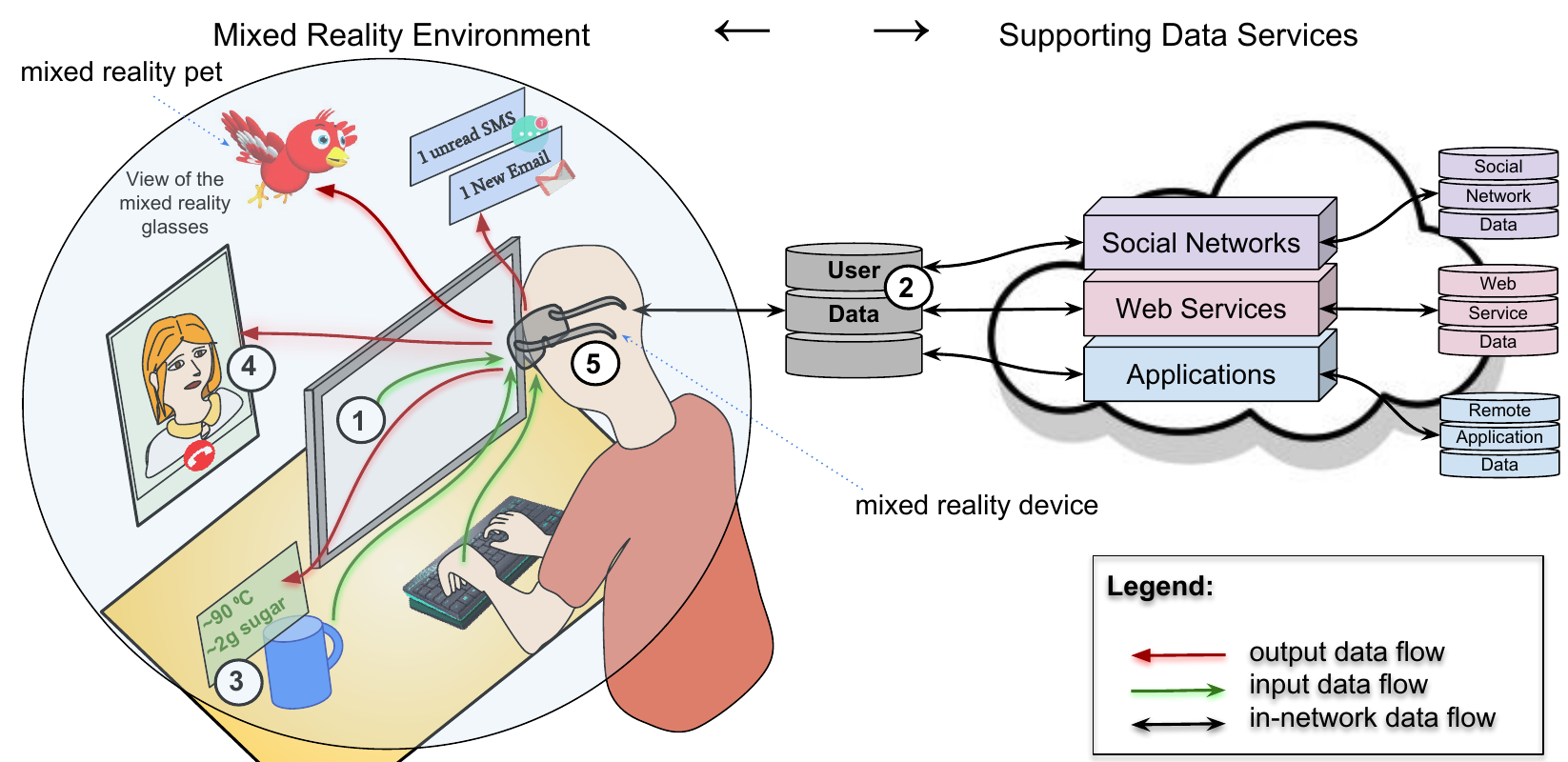}
\vspace{-4mm}
\caption{A mixed reality environment with the supporting data services as well as example points of protection as labelled: (1) contents of the display monitor, (2) access to stored data, (3) virtual display for content, e.g. information about the contents of a smart mug, (4) collaborating with other users, and (5) device access to the mixed reality eye-wear.}
\label{fig:mr-environment}
\vspace{-3mm}
\end{figure*}

Figure \ref{fig:mr-environment} presents an example of an MR environment with the supporting data services and shows how data flows \textit{within} the environment, and \textit{through} the data services. The left-half of the diagram shows the `view' of the mixed reality device which, in this example, is a see-through MR head-mounted device (HMD) or, simply, an MR eye-wear. Within the view are the physical objects which are `seen' by the MR device as indicated by the green arrows. The synthetic augmentations are shown in the diagram which are represented by the red arrows. The right-half of the diagram shows the various supporting data services that processes the data. The bi-directional solid arrows represent the access of these applications to captured data and the delivery of outputs to be augmented. Representative points of the five aspects of protection within the data flow are also labelled in Figure \ref{fig:mr-environment} and we use these labels to further explain the categorization as follows.

\begin{enumerate}
	\item \textit{Input Protection} -- This first category focuses on the challenges in ensuring security and privacy of data that is gathered and inputted to the MR platform. These data can contain sensitive information. For example, in Figure \ref{fig:mr-environment}, the MR eye-wear can capture the sensitive information on the user's desktop screen (labelled 1) such as e-mails, chat logs, and so on. These are user-sensitive information that needs to be protected. Similarly, the same device can also capture information that may not be sensitive to the user but may be sensitive to other entities such as bystanders. This is called \textit{bystander privacy}. Aside from readily sensitive objects, the device may capture other objects in the environment that are seemingly benign (or subtle) and were not intended to be shared but can be used by adversaries to infer knowledge about the users/bystanders. These necessary protections can be mapped to properties of \textit{confidentiality}, \textit{unobservability \& undetectability}, and \textit{content awareness}. The input protection approaches are discussed in \S\ref{sec:input}.
	
	\item \textit{Data Protection} -- After sensing, data is usually collected by the system in order for it to be processed. Data from multiple sensors or sources are aggregated, and, then, stored in a database or other forms of data storage. Applications, then, need to access these data in order to deliver output in the form of user-consumable information or services. However, almost all widely used computing platforms allows applications to collect and store data individually (as shown in the access of supporting data services labelled 2 in Figure \ref{fig:mr-environment}) and the users have no control over their data once it has been collected and stored by these applications. A lot of security and privacy risks have been raised concerning the access and use of user data by third party agents, particularly, on user data gathered from wearable \cite{felt2012android}, mobile \cite{lee2015risk}, and on-line activity \cite{ren2016recon}. MR technology faces even greater risks as richer information can be gathered using its highly sensitive sensors. For data protection, there are a lengthy list of properties that needs to be addressed such as integrity, availability, confidentiality, unlinkability, anonymity \& pseudonymity, and plausible deniability among others. \S\ref{sec:data} will present a discussion of the \textit{data protection} approaches.
	
	\item \textit{Output Protection} -- After processing the data, applications send outputs to the mixed reality device to be displayed or rendered. However, in MR, applications may inadvertently have access to outputs of other applications. If an untrusted application has access to other outputs, then it can potentially modify those outputs making them unreliable. For example, in the smart information (labelled 3) hovering over the cup in Figure \ref{fig:mr-environment}, malicious applications can modify the sugar level information. 
	Also, it is possible that one application's output is another's input which necessitates multiple application access to an output object. The integrity, availability, and policy compliance as well as reliability properties of these outputs has to be ensured. All these should be safely considered in \textit{output protection}. The approaches are discussed in \S\ref{sec:output}.
	
	\item \textit{User Interaction Protection} -- MR \textit{mixes} or includes the utilization of other sensing and display interfaces to allow immersive interactions. Examples of these are \textit{room-scale} interfaces\footnote{Theoriz studio designed and developed a room-scale MR demonstration (Link: http://www.theoriz.com/portfolio/mixed-reality-project/).} which allow multiple users to interact in the same MR space -- combining virtual and physical spaces, or the virtual ``tansportation'' of users in MR video-conferecing\footnote{Microsoft's Holoportation demonstrates virtual teleportation in real-time (Link: https://www.microsoft.com/en-us/research/project/holoportation-3/).} to allow them to seemingly co-exist in the same virtual space while being fully-aware of their physical space. Thus, we expand the coverage of protection to ensure protected sharing and collaborations (labelled 4 in Figure \ref{fig:mr-environment}) in MR.
	\subitem In contrast to current widely adapted technologies like computers and smart phones, MR can enable entirely new and different ways of interacting with the world, with machines, and with other users. One of the key expectations is how users can share MR experiences with assurance of security and privacy of information. Similar to data protection, there is a number of properties that is necessary in interaction protection namely non-repudiation, authorization, authentication, identifiability, and policy \& consent compliance. Details of the approaches in protecting user interactions are discussed in \S\ref{sec:interactions}.
	
	\item \textit{Device Protection} -- This last category focuses on the actual \textit{physical} MR device, and the physical input and ouput \textit{interfaces} of these devices. By extension, implicitly protects data that goes through all the other four aspects by ensuring device-level protection. Authentication, authorization, and identification are among the most important properties for device protection. In \S\ref{sec:device}, the different novel approaches in device access and physical display protection are discussed.
\end{enumerate}

\begin{figure*}[t]
	\centering
	\includegraphics[width=\textwidth]{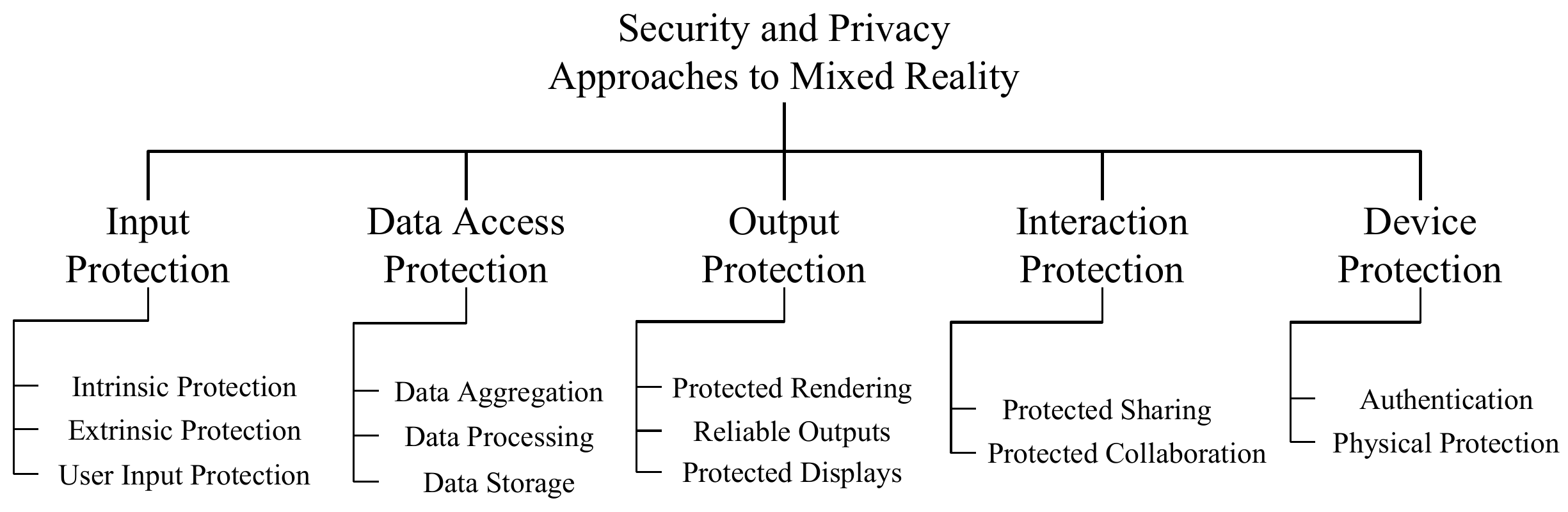}
	\vspace{-3mm}
	\caption{A data-centric categorization of the various security and privacy work or approaches on mixed reality and related technologies.}
	\label{fig:categories}
	\vspace{-3mm}
\end{figure*}
\begin{figure}[t]
	\centering
	\includegraphics[width=0.7\textwidth]{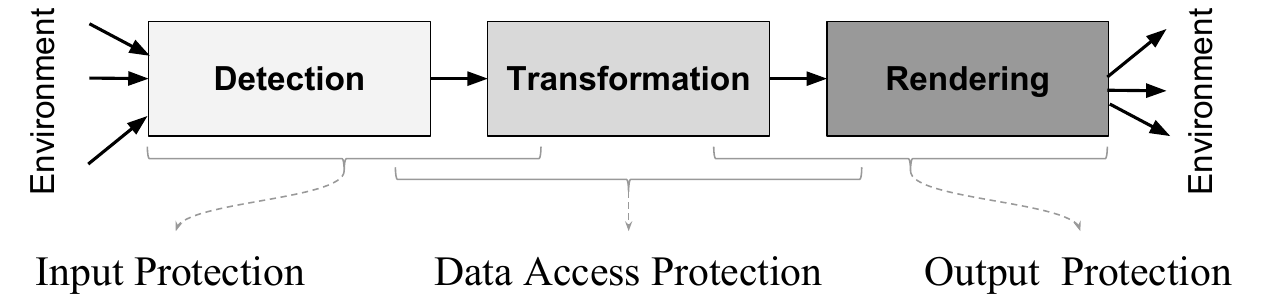}
	\vspace{-2mm}
	\caption{Generic mixed reality pipeline with the application target areas of three out of the five categories.}
	\label{fig:MRpipeline}
	\vspace{-3mm}
\end{figure}

Figure \ref{fig:categories} shows these categories with their further subcategories. A simplified representation of the process described in Figure \ref{fig:mr-environment} is shown as a pipeline in Figure \ref{fig:MRpipeline} which now has three essential blocks: \textit{detection}, \textit{transformation}, and \textit{rendering}. The first three categories are directly mapped to the associated risks with the main blocks of the processing pipeline -- protecting how applications, during the \textit{transformation} stage, access real-world input data gathered during \textit{detection}, which may be sensitive, and generate reliable outputs during \textit{rendering}. The detection focuses on gathering information such as user view orientation and direction, location, and surrounding objects. Thus, the detection process primarily depends on the sensing capabilities of the device. After detection, the information gathered will be transformed or processed to deliver services. Depending on the service or application, different transformations are used. Finally, the results of the transformation are delivered to the user by rendering it through the device's output interfaces. However, actual approaches may actually be applied beyond the simple boundaries we have defined here. Thus, some input and output protection approaches are actually applied in the transformation stage, while some data access protection approaches, e.g. data aggregation, are usually applied in the detection and rendering stages. Furthermore, the interaction protection and device protection approaches cannot be directly laid out along the pipeline unlike the other three as the intended targets of these two categories transcend this pipeline.

The presented categorization does not exclusively delineate the five aspects, and it is significant to note that most of the approaches that will be discussed can fall under \textit{more than one} category or subcategory. Notwithstanding, this survey paper complements the earlier surveys by presenting an up-to-date collection of security and privacy research and development for the past two decades on MR and related technologies and categorizing these various works according to the presented data-centric categorization. The next section proceeds in discussing these various approaches that have been done to address each of the five major aspects.

\section{Security and Privacy Approaches}\label{sec:s&p}

%

\begin{figure}[t]
		\centering
		\includegraphics[width=\textwidth]{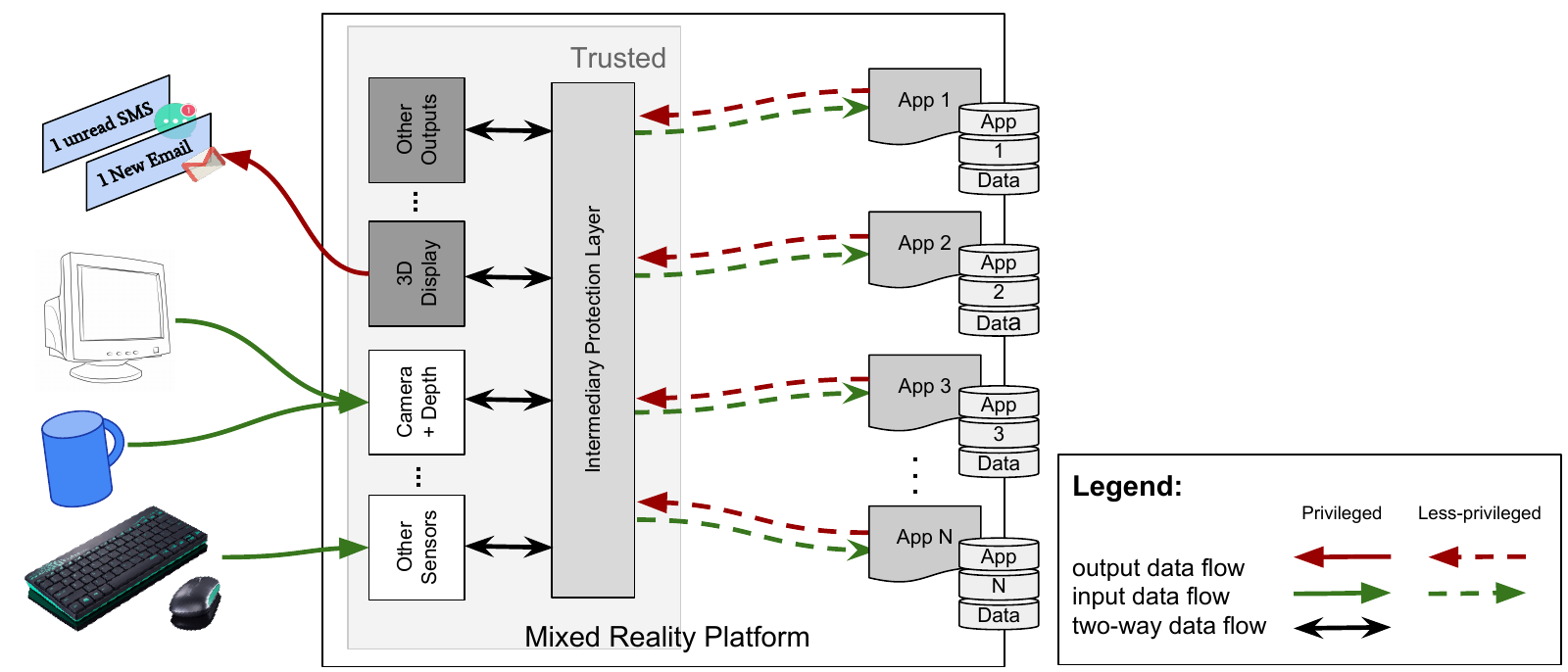}
		\caption{Shows a generic block diagram that inserts an \textit{\textbf{intermediary protection}} layer between the applications and device resources. (Data flows to and from third-party applications are now limited or less-privileged as represented by the broken arrows.)}\label{fig:intermediary_layer}
\vspace {-3mm}
\end{figure}

The MR environment diagram in Figure \ref{fig:mr-environment} shows how applications or third-party services access and utilize user data, and how once these applications have been granted access to those resources, they may now have indefinite access to them. Most file systems of popular mobile platforms (i.e. Android and iOS) have a \textit{dedicated} file system or database for each application. There are additional measures for security and privacy such as permission control, and application sand-boxing; however, despite these existing protection mechanisms, there are still pertinent threats that are not addressed. Thus, there is a great deal of effort to investigate data protection mechanisms. Most approaches rely on inserting an \textit{intermediary protection layer} (as shown in Figure \ref{fig:intermediary_layer}) that enables in- and out-flow control of data from trusted elements to untrusted ones{\let\thefootnote\relax\footnote{The rest of the figures in this survey paper will represent limited or \textit{less-privileged} data flow with broken arrow lines, while privileged ones are represented by complete arrows as shown in Fig.\ref{fig:intermediary_layer}'s legend.}}. 

In the following subsections, we present the various security and privacy work that has been done on MR and related technologies, especially on AR. We have organized these approaches according to the five major categories and, if applicable, to their further subcategories. There may be instances that presented solutions may address several aspects and may fall on more than one category. For these cases, we focus on the primary objective or challenges focused on their approach. We first discuss the threats specific to that category and, then, follow it with the security and privacy approaches, and strategies both in literature and those used in existing systems.

\subsection{Input Protection}\label{sec:input} Perhaps, the main threat to input protection, as well as to the other four categories, is the unauthorized and/or unintended disclosure of information -- may it be of actual data entities or of their flow. These vulnerable inputs can be categorized in to two based on the user intention. Targeted physical objects and other \textit{non}-user-intended inputs are both captured from the environment usually for visual augmentation anchoring. We can collectively call these inputs as \textit{passive}, while those that are intentionally provided by users, such as \textit{gestures}, can be considered as \textit{active inputs}. 

\subsubsection{Threats to \textbf{Passive Inputs: Targeted} and \textbf{Non-intended Latent Data}}\label{sssec:sanitization} Aside from threats to confidentiality (i.e. information disclosure), the two other main threats in the input side are \textit{detectability} and user \textit{content unawareness}. Both stems from the fact that these MR systems (just like any other service that employs a significant number of sensors) collects a lot of information, and among these are necessary and sensitive information alike. As more of these services becomes personalized, the sensitivity of these information increases. These threats are very evident with visual data. MR, as well as AR and VR, requires the detection of targets, i.e. objects or contexts, in the real environment, but other non-necessary and sensitive information are captured as well. These objects become detectable and users may not intend for these \textit{latent} information or contexts to be detected. Likewise, as they continuously use their MR device, users may not be made aware that the applications running on their device are also able to collect information about these objects and contexts.

\begin{figure}[t]
	\centering
	\includegraphics[width=0.5\textwidth]{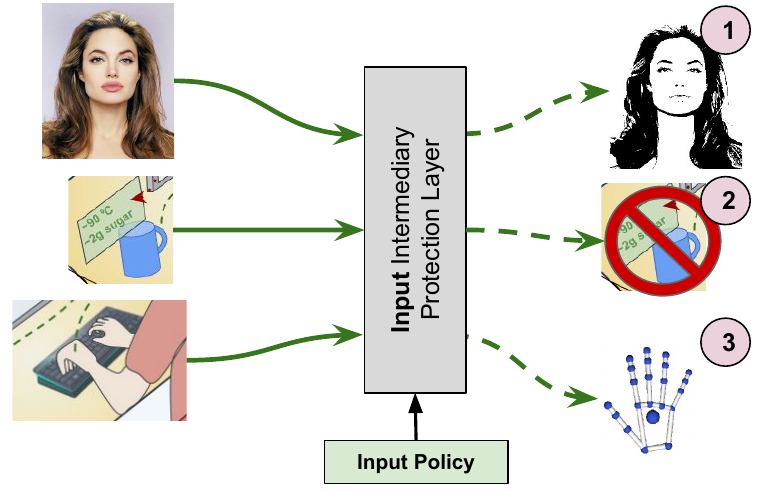}
	\caption{Example strategies for input protection: 1) \textit{information reduction} or \textit{partial sanitization}, e.g. from RGB facial information to facial outline only; 2) \textit{complete sanitization} or blocking; or 3) skeletal information instead of raw hand video capture.}\label{fig:input_protection_strategies}
	\vspace {-3mm}
\end{figure}

\paragraph{Protection Approaches} The most common input protection approaches usually involves the removal of latent and sensitive information from the input data stream. These approaches are generally called \textit{input sanitization} techniques (see samples labelled 1 and 2 in Figure \ref{fig:input_protection_strategies}). These are usually implemented as an intermediary layer between the sensor interfaces and the applications as shown in Figure \ref{fig:intermediary_layer}. In general, this protection layer acts as an input access control mechanism aside from sanitization. These techniques can further be categorized according to the policy enforcement -- whether \textit{intrinsic} or \textit{extrinsic} policies for protection are used. With intrinsic enforcement, the user, device, or system itself imposes the protection policies that dictates the input sanitization that is applied. On the other hand, extrinsic input protection arises from the need for sensitive objects \textit{external} to the user that are not considered by the intrinsic policies. In the following subsections, the sanitization techniques are presented as either intrinsic or extrinsic approaches. 

\begin{enumerate}
	\item \textit{Intrinsic input sanitization} policies are usually user-defined. For example, the \textsc{Darkly} system \cite{jana2013scanner} for perceptual applications uses OpenCV in its intermediary input protection layer to implement a multi-level feature sanitation. The basis for the level or degree of sanitization are the user-defined policies. The users can impose different degrees of sensitivity permissions which affects the amount of detail or features which can be provided to the applications, i.e. stricter policies mean less features are provided. For example, facial information can vary from showing facial feature contours (of eyes, nose, brows, mouth, and so on) to just the head contour depending on the user's preferences. The user can actively control the level of information that is provided to the applications. Thus, aside from providing \textit{undetectability \& unobservability}, and \textit{content awareness} to users, \textsc{Darkly} also provides a form of \textit{authorization} through information access control, specifically a \textit{least privilege} access control.
	
		\subitem \textit{Context-based Sanitization.} A context-based intrinsic sanitization framework \cite{zarepour2016context} improves on the non-contextual policies of \textsc{Darkly}. It determines if there are sensitive objects in the captured images, like faces or car registration plates, and automatically implements sanitization. Sensitive features are sanitized by blurring them out, while images of sensitive locations (e.g. bathrooms) are deleted entirely. Similarly, \textsc{PlaceAvoider} \cite{templeman2014placeavoider} also detects images as sensitive or not, depending on the features extracted from the image, but deletion is not automatic and still depends on the user. Despite the context-based nature of the sanitization, the policy that governs how to interpret the extracted contexts are still user-defined, thus, we consider both sanitization techniques as intrinsic. However, intrinsic policy enforcement can be considered as self-policing which can potentially have a \textit{myopic} view of privacy preferences of other users and objects. Furthermore, intrinsic policies can only protect the inputs that are explicitly identified in the policies. 
		
		\subitem \textit{Video Sanitization.} The previously discussed sanitization techniques were targeted for generic capturing devices and were mostly sanitizing images and performs the sanitization after the image is stored. For MR platforms that require real-time video feed, there is a need for live and on-the-fly sanitization of data to ensure security and privacy. A privacy-sensitive visual monitoring \cite{szczuko2014augmented} system was implemented by removing persons from a video surveillance feed and render 3D animated humanoids in place of the detected and visually-removed persons. Another privacy-aware live video analytic system called \textsc{OpenFace-RTFace} \cite{wang2017scalable} focused on performing fast video sanitization by combining it with face recognition. 
		The \textsc{OpenFace-RTFace} system lies near the \textit{edge} of the network, or on \textit{cloudlets}. Similar approaches to edge or cloud-assisted information sanitization can potentially be utilized for MR.
	
	\item \textit{Extrinsic input sanitization}\label{sssec:extrinsic} receives input policies, e.g. privacy preferences, from the environment. An early implementation \cite{truong2005preventing} involved outright capture interference to prevent sensitive objects from being captured by unauthorized visual capturing devices. A camera-projector set up is used. The camera detects unauthorized visual capture devices, and the projector beams a directed light source to ``blind'' the unauthorized device. This technique can be generalized as a form of a \textit{physical access control}, or, specifically, a deterrent to physical or visual access. However, this implementation requires a dedicated set up for every sensitive space or object, and the light beams can be disruptive to regular operation.
	
		\subitem Other approaches involves the use of existing communication channels or infrastructure for endorsing or communicating policies to capture devices, and to ensure that enforcement is less disruptive. The goal was to implement a fine-grained permission layer to ``automatically'' grant or deny access to continuous sensing or capture of any real-world object. A simple implementation on a privacy-aware see-through system \cite{hayashi2010installation} allowed other users that are ``seen-through'' to be blurred out or sanitized and shown as human icons only if the viewer is not their friend. However, this requires that users have access to the shared database and explicitly identify friends. Furthermore, enabling virtually anyone or, in this case, anything to specify policies opens new risks such as forgery, and malicious policies.
		
		\subitem To address authenticity issues in this so called \textit{world-driven} access control, policies can be transmitted as digital certificates \cite{roesner2014world} using a public key infrastructure (PKI). Thus, the PKI provides \textit{cryptographic} protection to \textit{media access} and sanitization policy transmission. However, the use of a shared database requires that all possible users' or sensitive objects' privacy preferences have to be pushed to this shared database. Furthermore, it excludes or, unintentionally, leaves out users or objects that are not part of the database --or, perhaps, are unaware -- which, then, defeats the purpose of a world-driven protection.
		
		\subitem \textsc{I-pic} \cite{aditya2016pic} removes the involvement of shared databases. Instead users endorse privacy choices via a peer-to-peer approach using Bluetooth Low Energy (BLE) devices. However, I-pic is only a capture-or-no system. \textsc{PrivacyCamera} \cite{li2016privacycamera} is another peer-to-peer approach but is not limited to BLE. Also, it performs face blurring, instead of just capture-or-no, using endorsed GPS information to determine if sensitive users are within camera view. On the other hand, \textsc{Cardea} \cite{shu2016cardea} allows users to use hand gestures to endorse privacy choices. In \textsc{Cardea}, users can show their palms to signal protection while a peace-sign to signal no need for protection. However, these three approaches are primarily targeted for bystander privacy protection, i.e. facial information sanitization.
		
		\subitem \textsc{MarkIt} \cite{raval2014markit} can provide protection to any user or object uses privacy markers and gestures (similar to \textsc{Cardea}) to endorse privacy preferences to cameras. It was integrated to Android's camera subsystem to prevent applications from leaking private information \cite{raval2016you} by sanitizing sensitive media. This is a step closer to automatic extrinsic input sanitization, but it requires visual markers in detecting sensitive objects. Furthermore, all these extrinsic approaches have only been targeted for visual capture applications and not with AR- or MR-specific ones.
	
\end{enumerate}

\subsubsection{Threats to \textbf{Gestures and other Active User Inputs}}\label{sssec:userInputs}
Another essential input that needs to be protected is \textit{gesture input}. We put a separate emphasis on this as gesture inputs entails a `direct' command to the system, while the previous latent and user inputs do not necessarily invoke commands. 
Currently, the most widely adopted user input interfaces are the tactile types, specifically, the keyboard, computer mouse, and touch interfaces. However, these current tactile inputs are limited by the dimension\footnote{Keyboards and other input pads can be considered as one-dimensional interfaces, while the mouse and the touch interfaces provides two-dimensional space interactions with limited third dimension using scroll, pan, and zoom capabilities.} of space that they are interacting with and some MR devices now don't have such interfaces. Also, these input interface types are prone to a more physical threat such as \textit{external inference} or \textit{shoulder-surfing} attack. From which, threats such as \textit{spoofing}, \textit{denial of service}, or \textit{tampering} may arise.

Furthermore, there is a necessity for new user input interfaces to allow three-dimensional inputs. Early approaches used gloves \cite{dorfmuller2001finger,thomas2002glove} that can determine hand movements, but advances in computer vision have led to tether- and glove-free 3D interactions. Gesture inference from smart watch movement have also been explored as a possible input channel, particularly on finger-writing inference \cite{xu2015finger}. Now, vision-based \textit{natural user interfaces} (NUI), such as the Leap Motion \cite{zhao2016leapmotion} and Microsoft Xbox Kinect, have long been integrated with MR systems to allow users to interact with virtual objects beyond two dimensions. This allows the use of body movement or gestures as input channels and move away from keypad and keyboards.  However, the use of visual capture to detect user gestures or using smart watch movement to detect keyboard strokes means that applications that require gesture inputs can inadvertently capture other sensitive inputs \cite{maiti2016smartwatch}. Similar latent privacy risks such as detectability and content unawareness arise. Thus, while new ways of interacting in MR are being explored, security and privacy should also be ensured.

\paragraph{Protection through abstraction} \textsc{Prepose} \cite{figueiredo2016prepose} provides secure gesture detection and recognition as an intermediary layer (as in Figure \ref{fig:intermediary_layer}). The \textsc{Prepose} core only sends gesture events to the applications, which effectively removes the necessity for untrusted applications to have access to the raw input feed. Similar to \textsc{Darkly}, it provides \textit{least privilege} access control to applications, i.e. only the necessary gesture event information is transmitted to the third party applications and not the raw gesture feed. 

A preceding work to \textsc{Prepose} implemented the similar idea of inserting a hierarchical recognizer \cite{jana2013enabling} as an intermediary input protection layer. They inserted \textsc{Recognizers} to the Xbox Kinect to address input sanitization as well as to provide input access control. The policy is user-defined; thus, it is an intrinsic approach. Similarly, the goal is to implement a \textit{least privilege} approach to application access to inputs -- applications are only given the least amount of information necessary to run. For example, a dance game in Xbox, e.g. Dance Central or Just Dance, only needs body skeletal (similar to sample labelled 3 in Figure \ref{fig:input_protection_strategies}) movement information, and it does not need facial information, thus, the dance games are only provided with the moving skeletal information and not the raw video feed of the user while playing. 
To handle multiple levels of input policies, the recognizer implements a hierarchy of privileges in a tree structure, with the root having highest privilege, i.e. access to RGB and depth information, and the leaves having lesser privileges, i.e. access to skeletal information.

\subsubsection{Remaining Challenges in Input Protection} Most of the approaches to input protection are founded on the idea of \textit{least privilege}. However, it requires that the intermediary layer, e.g. the \textsc{Recognizers}, must know what type of inputs or objects the different applications will require. \textsc{Prepose} addresses this for future gestures but not for future objects. For example, an MR painting application may require the detection of different types of brushes but the current recognizer does not know how to `see' or detect the brushes. Extrinsic approaches like \textsc{MarkIt} try to address this by using markers to tell which objects can and cannot be seen. What seemingly arises now is the need to have a \textit{dynamic} abstraction and/or sanitization of both pre-determined and future sensitive objects. 
In \S\ref{sssec:interface_protection}, we will focus on device-level protection approaches to protect user activity involving input interfaces.

\subsection{Data Protection}\label{sec:data}

We can divide the different data protection techniques based on the data flow. First, after sensing, data is gathered and aggregated, thus, protected \textit{data aggregation} is necessary. Then, to deliver output, applications will need to process the data, thus, privacy-preserving \textit{data processing} is required. Ultimately, the \textit{data storage} has to be protected as well. Thus, three major data protection layers arise: aggregation, processing, and storage.

Generally, the aim of these data protection approaches is to allow services or third-party applications to learn something without leaking unnecessary and/or personally identifiable information. Usually, these protection approaches use privacy definitions such as \textit{k}-anonymity, and \textit{differential privacy}. \textit{k}-anonymity \cite{samarati2001protecting, samarati1998protecting} ensures that records are unidentifiable from at least \textit{k-1} other records. It usually involves data perturbation or manipulation techniques to ensure privacy, but suffers from scaling problems, i.e. larger data dimensions, which can be expected from MR platforms and devices with a high number of sensors or input data sources. Differentially private algorithms \cite{dwork2014algorithmic}, on the other hand, inserts randomness to data to provide \textit{plausible deniability} and \textit{unlinkability}. The guaranteed privacy of differentially private algorithms is well-studied \cite{mcsherry2007mechanism}. Ultimately, there are other privacy definitions or metrics used in the literature. In the following subsections, we now focus on the different threats and approaches to data aggregation, processing, and storage.

\subsubsection{Threats to \textbf{Data Collection and Aggregation}}\label{ssec:data_collection}
Essentially, data collection also falls under the input category but we transfer the focus to data \textit{after} sensing and how systems, applications, and services handle data as a whole. The main threats to data collection are \textit{tampering}, \textit{denial of service}, and \textit{unauthorized access} among others. These are primarily security threats. For example, an adversary can tamper MR targets to elicit a different response from the system or to outright deny a service. For these types of threats, we will focus on these later under \textit{device protection} on \S\ref{sec:device}.

Aside from security threats, \textit{linkability}, \textit{detectability}, and \textit{identifiability} are some of the privacy threats that results from continuous or persistent collection of data. Aggregation further aggravates these threats due to the increase in the number of channels and the amount of information.

\subsubsection*{Protected Data Collection and Aggregation}Protected data collection and aggregation approaches are also implemented as an intermediate layer as in Figure \ref{fig:intermediary_layer}. Usually, data perturbation or similar mechanisms are run on this intermediary layer to provide a privacy guarantee, e.g. differential privacy or k-anonymity.

\begin{enumerate}
	\item \textit{Privacy-preserving data collection and aggregation.} RAPPOR or \textit{randomized response} \cite{erlingsson2014rappor} is an example of a differentially-private data collection and aggregation algorithm. It is primarily applied for privacy-preserving crowd-sourced information such as those collected by Google for their Maps services. Privacy-preserving data aggregation (PDA) has also been adopted for information collection systems \cite{he2007pda, he2011pda} with multiple data collection or sensor points, such as wireless sensor networks or body area networks. The premise of PDA is to get aggregate statistic or information without knowing individual information, whether of individual sensors or users. A similar PDA approach specific to MR data is still yet to be designed, developed, and evaluated.
	
	\item \textit{Using abstraction for authorized sensor access.} \textsc{SemaDroid} \cite{xu2015semadroid}, on the other hand, is a device level protection approach. It is a privacy-aware sensor management framework that extends the current sensor management framework of Android and allows users to specify and control fine-grained permissions to applications accessing sensors. Just like the abstraction strategies in input protection, \textsc{SemaDroid} is implemented as an intermediary protection layer that provides users application access control or \textit{authorization} to sensors and sensor data. What differentiates it from a the input protection techniques are the application of \textit{auditing} and \textit{reporting} of potential leakage and applying them to a \textit{privacy bargain}. This allows users to `trade' their data or privacy in exchange for services from the applications. There are a significant number of work on privacy bargain and the larger area of \textit{privacy economics} but we will not be elaborating on it further and point the readers to Acquisti's work on privacy economics \cite{acquisti2016economics}.
\end{enumerate}

\subsubsection{Threats to \textbf{Data Processing}}\label{ssec:data_processing}
After collection, most services will have to process the data immediately to deliver outputs fast and real-time. Similar to data collection, the same privacy threats of \textit{information disclosure}, \textit{linkability}, \textit{detectability}, and \textit{identifiability} holds. During processing, third-party applications or services can directly access user data which may contain sensitive or personal information if no protection measures are implemented.

\subsubsection*{Protection Approaches} The same premise holds: applications process data to deliver services but with data security and user privacy in mind. Both secure and privacy-preserving data processing algorithms can be applied to achieve both security and privacy properties such as \textit{integrity}, \textit{confidentiality}, \textit{unlinkability}, and \textit{plausible deniability} among others.

\begin{figure}[t]
	\centering
	\includegraphics[width=0.9\textwidth]{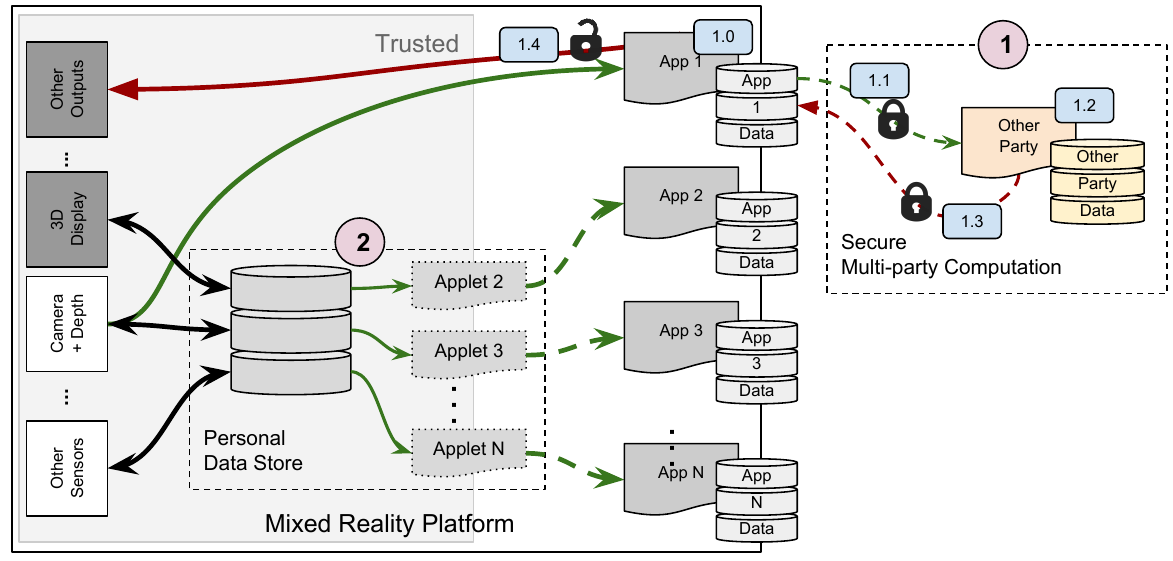}
	\caption{Generic block diagrams of two different data protection approaches: 1) cryptographic technique using \textit{secure multi-party} computation where two or more parties exchange secrets (1.1 and 1.3) to extract combined knowledge (1.2 and 1.4) without the need for divulging or decrypting each others data share; and 2) personal data stores with ``trusted'' applets.}
	\label{fig:data_protection}
	\vspace {-3mm}
\end{figure}

\begin{enumerate}
	\item \textit{Encryption-based techniques.} Homomorphic encryption (HE) allows queries or computations over encrypted data. There are varying levels of homomorphism from partial, such as Paillier encryption \cite{paillier1999public} which is homomorphic in addition (and, to some extent, multiplication), to \textit{fully homomorphic encryption} (FHE) \cite{gentry2009fully}. Upon collection, data is encrypted first and homomorphism allows third-party service to make computations over data without decryption. This technique is primarily used for remote data processing especially when data processors are not trusted.
	
	\subitem In visual data processing, encryption-based techniques have been used for image feature extraction, and matching for various uses such as image search, and context or object detection. \textsc{He-Sift} \cite{hsu2011homomorphic} performs bit-reversing and local encryption to the raw image before feature description using SIFT\footnote{SIFT or \textit{Scale-invariant Feature Transform} is a popular image feature extraction and description algorithm} \cite{lowe2004distinctive}. The goal was to make dominant features, which can be used for context inference, recessive. As a result, feature extraction, description, and matching are all performed in the encrypted domain. A major drawback is the very slow computation time due to the near full-homomorphism used as well as the approach being algorithm-specific. Using \textit{leveled} HE can reduce the computation time of \textsc{He-Sift} \cite{jiang2017secure}.
	
	\subitem Other improvements utilizes big data computation techniques to expedite secure image processing such as the use of a combination of MapReduce and \textit{ciphertext-policy attribute-based encryption} \cite{zhang2014cloud}, or the use of Google's Encrypted BigQuery Client for Paillier HE computations \cite{ziad2016cryptoimg}.
	
	\item \textit{Secret Sharing or Secure Multi-party Computation.} Data can be split among untrusted parties assuming that information can only be inferred when the distributed parts are together. \textit{Secure multi-party computation} (SMC) or \textit{secret sharing} allows computation of data from two or more sources without necessarily knowing about the actual data each source has. The diagram labelled 1 in Figure \ref{fig:data_protection} shows a possible SMC setup. For example, App 1 requires data from another party (could be anohter application) to provide a certain service. It encrypts its share of the data (step 1.0) and sends it to the other party (1.1). The other party then encrypts its other share (1.2) and sends it to App 1 (1.3). Both can compute the results over the combined encrypted shares without the need to decrypt their shares.
	
	\subitem There are various approaches in performing secret sharing such as garbled circuits \cite{yao1986generate, huang2011faster}, and other non-cryptographic-based ones. For example, \textsc{SecSift} \cite{qin2014private,qin2014towards} improves on the computation time of \textsc{He-Sift} by using a \textit{somewhat homomorphic encryption}. They split or distribute the SIFT feature computation tasks among a set of ``independent, co-operative cloud servers [to] keep the outsourced computation procedures as simple as possible [and] avoid utilizing homomorphic encryption.'' There are two primary image processing tasks that SecSIFT provide \cite{qin2014privacy}: color histogram and layout descriptors. These two tasks allows for a variety of image processing tasks compared to \textsc{He-Sift} which is specific to feature matching.
	
	\subitem P3 \cite{ra2013p3} focuses on photo sharing and uses two-party secret sharing by ``by splitting a photo into a public part, which contains most of the volume (in bytes) of the original, and a secret part which contains most of the original's information.'' It uses \textit{AES-based symmetric keys} to encrypt the secret part and allows the use of a tunable parameter between storage/bandwidth and privacy. This approach, however, is JPEG-format specific.
	
	\subitem A privacy-preserving virtual cloth try-on \cite{sekhavat2017privacy} service used secret sharing and secure two-party computation. The anthropometric information\footnote{Anthropometric information are body measurements of an individual that capture size, shape, body composition, and so on.} of the user is split between the user's mobile device and the server, and are both encrypted. The server has a database of clothing information. The server can then compute a 3D model of the user wearing the piece of clothing by combining the anthropometric information and the clothing information to generate an encrypted output which is sent to the user device. The user device decrypts the result and combines it with the local secret to reveal the 3D model of the user ``wearing'' the piece of clothing.
	
	\item \textit{Virtual Reconstructions.} We can take advantage of the \textit{artificiality} that MR provides. Instead of providing complete 3D (RGB+depth) data, a 3D sanitized or `salted' virtual reconstruction of the physical space can be provided to third-party applications. For example, instead of showing the 3D capture of a table in the scene with all 3D data of the objects on the table, a generalized horizontal platform or surface can be provided. The potentially sensitive objects on the table are kept confidential. A tunable parameter balances between sanitization (or salting) and reconstruction latency. Using this tunability, similar notions of privacy guarantee can be provided such as differential privacy and k-anonymity. However, this approach is yet to be realized but virtual reconstruction has been used to address delayed alignment issues in AR \cite{waegel2014poster}. This approach can work well with other detection (\S\ref{sssec:userInputs}) and rendering (\S\ref{sssec:secureRendering}) strategies of sanitization and abstraction as well as in privacy-centred collaborative interactions (\S\ref{sssec:collabInteractions}). This approach also opens the possibility to have an \textit{active defence} strategy where 'salted' reconstructions are offered as a honeypot to adversaries.

\end{enumerate}

All these techniques complement each other and can be used simultaneously on a singular system. Inevitably, we understand the equal importance of all these technologies and how they can be used on MR, but these data protection techniques are technology agnostic. Therefore, any or all of these techniques can be applied to MR and it will only be a matter of whether the technique is appropriate for the amount of data and level of sensitivity of data that is tackled in MR environments.

\subsubsection{Threats to \textbf{Data Storage}}\label{ssec:data_storage}
After collection and aggregation, applications store user data on separate databases in which users have minimal or no control over. Privacy concerns on how these applications use user data beyond the expected utility to the user have been posed \cite{ren2016recon, lee2015risk, felt2012android}. Aside from these privacy threats, there are inherent security threats such as \textit{tampering}, \textit{unauthorized access}, and \textit{spoofing} that are faced by data storage.

\subsubsection*{Data Storage Solutions to Protection} When trustworthiness is not ensured, protected data storage solutions, such as \textit{personal data stores} (PDS), with managed application access permission control is necessary. PDSs allows the users to have control over their data, and which applications have access to it. In addition, further boundary protection and monitoring can be enforced on the flow of data in and out of the PDS. Figure \ref{fig:data_protection} shows a generic block diagram (labelled 2) of how a PDS protects the user data by running it in a protected sand-box machine that may monitor the data that is provided to the applications. Usually, applet versions (represented by the smaller App blocks within the PDS) of the applications run within the sand-box. Various PDS implementations have been proposed such as the \textit{personal data vaults} (PDV), \textsc{OpenPDS}, and the \textsc{Databox}.

\begin{enumerate}
\item The PDV was one of the earlier implementations of a PDS but it only supports a few number of data sources, i.e. location information, and does not have an \textit{application programming interface} (or API) for other data sources or types. Nonetheless, they demonstrated how data storage and data sharing to applications can be decoupled \cite{mun2010pdv}.

\item \textsc{OpenPDS} improves on the lack of API of the PDV. \textsc{OpenPDS} allows any application to have access to user data through \textsc{SafeAnswers} \cite{deMontjoye2014openpds}. \textsc{SafeAnswers} (SA) are pre-submitted and pre-approved query application modules (just like an applet in Figure \ref{fig:data_protection}) which allows applications to retrieve results from the PDS using. However, the necessity of requiring applications to have a set of pre-approved queries reduces the flexibility of openPDS.

\item \textsc{Databox} also involves the use of a sandbox machine where users store their data, and applications run a containerized piece of query application that is trusted. The running of containers in the sandbox allows users to have control of what pieces of information can exit from the sandbox. However, similar to \textsc{SafeAnswers} requiring application developers to develop a containerized application for their queries may be a hindrance to adaptation. Despite that, \textsc{Databox} pushes for a \textit{privacy ecosystem} which empowers users to trade their data for services similar to \textit{privacy bargains} in a \textit{privacy economy} \cite{crabtree2016databox}. This privacy ecosystem can possibly assist in addressing the adaptability issues of developing containerized privacy-aware query applications, because users can now demand service in exchange for their data.

\end{enumerate}

\subsubsection{Remaining Challenges in Data Protection}
Most of the data protection techniques discussed were implemented on generic systems and not necessarily MR-targeted. MR data is expected to be visual-heavy, and information collected is not only confined to users but also of other externally sensitive information that can be captured. Moreover, there are necessary modifications that applications have to partake in order to implement these data protection strategies. Aside from implementation complexity are the additional resources necessary such as the inherent need of memory, and compute capacity to use encryption-based schemes. There are attempts to eliminate the necessity of code modification such as \textsc{GUPT} \cite{mohan2012gupt}, which focuses on the sampling and aggregation process to ensure distribution of the differential privacy budget and eliminating the need for costly encryption. Also, combining these techniques with protected sensor management and data storage to provide \textit{confidentiality} through sanitization and \textit{authorized} access control is promising.

\subsection{Output Protection}\label{sec:output}
The prime value of MR is to deliver immersive experiences. To achieve that, applications ship services and experiences in the form of rendered outputs. In general, there are three possible types of outputs in MR systems: real-world anchored outputs, non-anchored outputs, and outputs of external displays. The first two types are both augmented outputs. The last type refers to outputs of other external displays which can be utilized by MR systems, and vice versa. Protecting these outputs is of paramount importance aside from ensuring input and data protection. As a result, there are three enduring points or aspects of protection when it comes to the output: protecting external displays, output control, and protected rendering.

\subsubsection{Threats to \textbf{Output Reliability} and \textbf{User Safety}}\label{sssec:safeOutputs} Current `reality' systems have loose output access control. As a result, adversaries can potentially \textit{tamper} or \textit{spoof} outputs that can compromise user safety. In addition, reliability is also compromised resulting to threats such as \textit{denial of service}, and \textit{policy \& consent non-compliance}.

\subsubsection*{Safe and Reliable Outputs} Output control policies can be used as a guiding framework on how MR devices will handle outputs from third-party applications. This includes the management of rendering priority which could be in terms of synthetic object transparency, arrangement, occlusion, and other possible spatial attributes. An output access control framework \cite{lebeck2016safely} with an object-level of granularity have been proposed to make output handling enforcement easier. It can be implemented as an intermediary layer, as in Figure \ref{fig:intermediary_layer}, and follows a set of output policies. In a follow up work, they presented a design framework \cite{lebeck2017securing} for output policy specification and enforcement which combined output policies from Microsft's HoloLens Developer guidelines, and the U.S. Department of Transportation's National Highway Traffic Safety Administration (NHTSA) (for user safety in automobile-installed AR). Here are two example descriptions of their policies: ``Don't obscure pedestrians or road signs" is inspired from the NHTSA; ``Don't allow AR objects to occlude other AR objects" is inspired from the HoloLens's guidelines. They designed a prototype platform called \textsc{Arya} that will implement the application output control based on the output policies specified and evaluated \textsc{Arya} on various simulated scenarios. As of yet, \textsc{Arya} is the only AR or MR output access control approach in the literature.

\subsubsection{Threats during \textbf{Rendering}}\label{sssec:secureRendering}
Other MR environments incorporates any surface or medium as a possible output display medium. For example, when a wall is used as a display surface in an MR environment, the applications that use it can potentially capture the objects or other latent and/or sensitive information within the wall during the detection process. This specific case intersects very well with the input category because what is compromised here is the sensitive information that can be captured in trying to determine the possible surfaces for displaying.

\subsubsection*{Privacy-preserving Rendering} Applications that requires such displays do not need to know what the contents in the wall are. It only has to know that there is a surface that can be used as a display. Protected output rendering protects the medium and, by extension, whatever is in the medium. \textit{Least privilege} has been used in this context \cite{vilk2014least}. For example, in a room-scale MR environment, only the skeletal information of the room, and the location and orientation of the detected surfaces (or display devices) is made known to the applications that wish to display content on these display surfaces \cite{vilk2015surroundweb}. This utilizes the same \textit{abstraction strategy} which have been used both in input and data protection to provide \textit{unobservability \& undetectability} and \textit{authorized} access. This example of room-scale MR environments is usually used for collaborative purposes. 

\subsubsection{Threats to \textbf{Output Displays}}\label{sssec:secretDisplays}
Output displays are vulnerable to physical inference threats or \textit{visual channel} exploits such as shoulder-surfing attacks. These are the same threats to user inputs (\S\ref{sssec:userInputs}) especially when the input and output interfaces are on the same medium or are integrated together such as touch screens.

\subsubsection*{Protecting outputs from external inference} To provide secrecy and privacy on certain sensitive contexts which requires output confidentiality (e.g. ATM bank transactions), MR can be leveraged to provide this kind of protection. This time, MR capabilities are leveraged for output defense strategies.

\begin{enumerate}
	\item \textit{Content hiding methods.} \textsc{EyeGuide} \cite{eaddy2004my} used a near-eye HMD to provide a navigation service that delivers secret and private navigation information augmented on a public map display. Because the \textsc{EyeGuide} display is practically secret, shoulder surfing is prevented.
	
	\subitem Other approaches involve the actual hiding of content. For example, \textsc{VRCodes} \cite{woo2012vrcodes} takes advantage of rolling shutter to hide codes from human eyes but can be detected by cameras at a specific frame rate. A similar approach has been used to hide AR tags in video \cite{lin2017video}. This type of technique can hide content from human attackers but is still vulnerable to machine-aided inference or capture.
	
	\item \textit{Visual cryptography.} Secret display approaches have also been used in visual cryptographic techniques such as visual secret sharing (VSS) schemes. VSS allows the `mechanical' decryption of secrets by overlaying the visual cipher with the visual key. However, classical VSS was targeted for printed content \cite{chang2010two} and requires strict alignment which is difficult even for AR and MR displays, particularly handhelds and HMDs. The VSS technique can then be relaxed by using code-based secret sharing, e.g. barcodes, QR codes, 2D barcodes, and so on. The ciphers are publicly viewable while the key is kept secret. An AR-device can then be used to read the cipher and augment the decrypted content over the cipher. This type of visual cryptography have been applied to both print \cite{simkin2014ubic} and electronic displays \cite{lantz2015visual, andrabi2015usability}.
	
	\subitem Electronic displays are, however, prone to attacks from malicious applications which has access to the display. One of these possible attacks is cipher rearrangement for multiple ciphers. To prevent such in untrusted electronic displays, a visual \textit{ordinal cue} \cite{fang2010securing} can be combined with the ciphers to provide the users immediate signal if they have been rearranged.
	
	\subitem These techniques can also be used to provide protection for sensitive content on displays during input sensing. Instead of providing privacy protection through post-capture sanitization, the captured ciphers will remain secure as long as the secret shares or keys are kept secure. Thus, even if the ciphers are captured during input sensing, the content stays secure. In general, these visual cryptography and content-hiding methods provide visual access control, and information protection in shared or public resources. More device-level examples of this technique are discussed in \S\ref{sssec:interface_protection}.
\end{enumerate}

\subsubsection{Remaining Challenges in Output Protection} Similar to input protection, output protection strategies can use the same \textit{abstraction} approach applied as an intermediary access control layer between applications and output interfaces or rendering resources. To enforce these output abstractions, a reference policy framework has to exist through which the abstraction is based upon. As a result, perhaps, the biggest challenge is the \textit{specification} and \textit{enforcement} of these policies -- particularly on who will specify them and how will they be effectively enforced. In the output side, risks and dangers are more imminent because adversaries are about to actuate or have already actuated the malicious response or output. Thus, these access control strategies and policies are significant approaches to output protection.

Likewise, malicious inference or capture of outputs present the same threats as input inference. \S\ref{sssec:interface_protection} will focus on device-level protection approaches to output interfaces and displays.

\subsection{Protecting User Interactions}\label{sec:interactions}

\begin{figure}[t]
	\begin{minipage}[t]{0.315\textwidth}
		\centering
		\includegraphics[width=\textwidth]{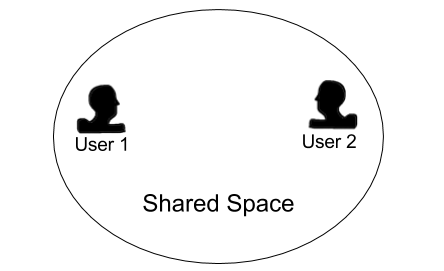}
		\subcaption{A simplified virtual \textit{shared space} diagram}\label{fig:shared_space_a}
	\end{minipage}
	\hfill
	\begin{minipage}[t]{0.315\textwidth}
		\centering
		\includegraphics[width=\textwidth]{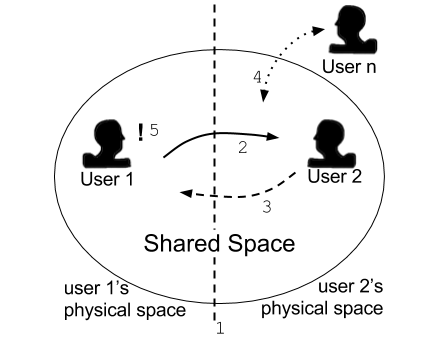}
		\subcaption{A possible separation in the underlying physical space which creates boundaries between users, and devices.}\label{fig:shared_space_b}
	\end{minipage}
	\hfill
	\begin{minipage}[t]{0.315\textwidth}
		\centering
		\includegraphics[width=\textwidth]{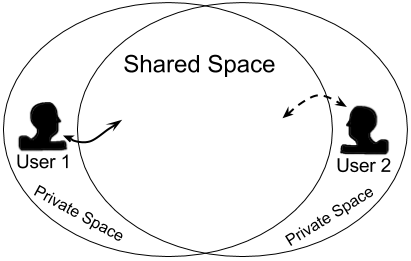}
		\subcaption{A collaborative space with both a shared space and a \textit{private space}.}\label{fig:shared_space_c}
	\end{minipage}
	\caption{Shared Spaces}
	\label{fig:shared_space}
	\vspace {-3mm}
\end{figure}

When it comes to collaboration, technologies such as audio-visual teleconferencing, and \textit{computer-supported collaborative work} (or CSCW) have been around to enable live sharing of information among multiple users. These are called \textit{shared space} technologies as more than one user interacts in the same shared space as shown in Figure \ref{fig:shared_space_a}. And MR offers a much more immersive sharing space. 

Concerns on the \textit{boundaries} -- ``transparent boundaries between physical and synthetic spaces'' -- in MR  and on the \textit{directionality} of these boundaries have been raised \cite{benford1998boundaries}. The directionality can influence the balance of power, mutuality and privacy between users in shared spaces. For example, the boundary (labelled 1) in Figure \ref{fig:shared_space_b} allows User 2 to receive full information (solid arrow labelled 2) from User 1 while User 1 receives partial information (broken arrow labelled 3) from User 2. The boundary enables an `imbalance of power' which can have potential privacy and ethical effects on the users.

Early attempts on ensuring user privacy in a collaborative MR context was to provide users with a private space, as shown in Figure \ref{fig:shared_space_c}, that displays outputs only for that specific user while having a shared space for collaboration. For example, the \textsc{PIP} or \textit{personal interaction panel} was designed to serve as a private interface for actions and tasks that the user does not want to share to the collaborative virtual space \cite{szalavari1997personal}. It is composed of a tracked ``dumb" panel and a pen. It was used as a gaming console to evaluate its performance. In the following subsection, we first take a look at some other early work on collaborative interactions, and, then, proceed with examples of how to manage privacy in shared spaces.

\subsubsection{Threats during \textbf{Collaborative Interactions}}\label{sssec:collabInteractions}
Early collaborative platform prototypes \cite{billinghurst1999collaborative, regenbrecht2002magicmeeting, grasset2002mare, schmalstieg2002, hua2004scape} demonstrated fully three-dimensional collaboration in MR. However, none have addressed the concerns raised from information sharing due to the boundaries created by shared spaces. An adversarial user can potentially \textit{tamper}, \textit{spoof}, or \textit{repudiate} malicious actions during these interactions. As a result, legitimate users may suffer \textit{denial of service} and may be \textit{unaware} that their personal data may have been captured and, then, leaked.

\subsubsection*{Protecting Collaborative Interactions} To provide protection during interactions, policy specification from users and its enforcement are the primary strategies. The system through which these interactions are coursed through should support the specification and enforcement of these policies.

\begin{enumerate}
	\item \textit{Enabling user-originated policies.} \textsc{Emmie} (Environmental Management for Multi-user Information Environments) \cite{butz1999emmie} is a \textit{hybrid} multi-interface collaborative environment which uses AR as a 3D interface at the same time allowing users to specify privacy of certain information or objects through \textit{privacy lamps} and \textit{vampire mirrors} \cite{butz1998vampire}. \textsc{EMMIE}'s \textit{privacy lamps} are virtual lamps that `emit' a light cone in which users can put objects within the light cone to mark these objects as private. On the other hand, the \textit{vampire mirrors} are used to determine privacy of objects by showing full reflections of public objects while private objects are either invisible or transparent. However, the privacy lamps and vampire mirrors only protect virtual or synthetic content and does not provide protection to real-world objects.

	\subitem \textsc{Kinected Conference} \cite{devincenzi2011kinected} allows the participants to use gestures to impose a temporary private session during a video conference. Aside from that, they implemented synthetic focusing using Microsoft Kinect's depth sensing capability -- other participants are blurred in order to direct focus on a participant who is speaking --, and augmented graphics hovering above the user's heads to show their information such as name, shared documents, and speaking time. The augmented graphics serve as \textit{feed-through} information to deliver signals that would have been available in a shared physical space but is not readily cross-conveyed between remote physical spaces.

	\item \textit{Feed-through signalling.} \textsc{SecSpace} \cite{reilly2014secspace} explores a feed-through mechanism to allow a more natural approach to user management of privacy in a collaborative MR environment. In contrast to \textsc{Kinected Conference}'s gesture-based privacy session, and \textsc{Emmie}'s privacy lamps and vampire mirrors, users in \textsc{SecSpace} are provided feed-through information that would allow them to negotiate their privacy preferences. Figure \ref{fig:shared_space_b} shows an example situation in which User n enters the shared space (labelled 4) on the same physical space as User 2 which triggers an alarm (labelled 5) or notification for User 1. The notification serves as a feed-through signalling that crosses over the MR boundary. By informing participants of such information, an \textit{imbalance of power} can be rebalanced through negotiations.
	
	\subitem Non-AR Feed-through signalling have also been used in a non-shared space context like the candid interactions \cite{ens2015candid} which uses wearable bands that lights up in different colors depending on the smart-phone activity of the user, or other wearable icons that change shape, again, depending on which application the icon is associated to. However, the pervasive nature of these feed-through mechanisms can still pose security and privacy risks, thus, these mechanisms should be regulated and properly managed. In addition, the necessary infrastructure, especially for \textsc{SecSpace}, to enable this pervasive feed-through system may be a detriment to adaptation. A careful balance between the users' privacy in a shared space and the utility of the space as a communication medium is ought to be sought.
	
	\item \textit{Private and public space interactions.} Competitive gaming demands secrecy and privacy in order to make strategies while performing other tasks in a shared environment. Thus, it is a very apt use case for implementing user protection in a shared space. \textsc{Private Interaction Panels} (or PIPs) demonstrates a gaming console functionality where a region that is defined within the \textsc{PIP} panel serves as a private region \cite{szalavari1998gaming}. For example, in a game of Mah-jongg, the \textsc{PIP} panel serves as the user's space for secret tiles while all users can see the public tiles through their HMDs. The \textsc{PIP} pen is used to pick-and-drop tiles between the private space and the public space. On the other hand, \textsc{TouchSpace} implements a larger room-scale MR game. It uses an HMD that can switch between see-through AR and full VR, an entire floor as shared game space with markers, and a wand for user interactions with virtual objects \cite{cheok2002touch}. Essentially, \textsc{Emmie}'s privacy lamps and mirrors also act as private spaces.
	
	\subitem \textsc{BragFish} \cite{xu2008bragfish} implements a similar idea on privacy to that of the \textsc{PIP} with the use of a handheld AR device, i.e. Gizmondo. In \textsc{BragFish}, a game table with markers serves as the shared space, while each user has the handheld AR that serves as the private space for each user. The handheld AR device has a camera that is used to ``read" the markers associated to a certain game setting, and it frees the user from the bulky HMDs as in \textsc{PIP} and \textsc{TouchSpace}. The Gizmondo handheld device has also been used in another room-scale AR game \cite{mulloni2008gameplay}. Similarly, camera phones have been used as a handheld AR device in a table top marker-based setup for collaborative gaming \cite{henrysson2005facetoface}.

\end{enumerate}

In collaborative mixed reality, aside from policy enforcement and feed-through signalling, there are two basic spaces that can be utilised during interactions: a shared space for public objects, and a private space for user-sensitive tasks such as making strategies. All examples, such as \textsc{PIP} and \textsc{BragFish}, assumes that each user can freely share and exchange content or information through a shared platform or interaction channel. Furthermore, these privacy-sensitive shared space approaches are also, to some extent, inherently distributed which provides further security and privacy. In the next section, we focus on how this simple act of sharing can be protected in a mixed reality context without a pre-existing shared channel.

\subsubsection{Threats to \textbf{Sharing Initialization}}\label{sssec:sharing}
All those shared space systems that were previously discussed rely on a unified architecture to enable interactions and sharing on the same channel. However, there might be cases that sharing is necessary but no pre-exisiting channel exists, or an entire architecture, just like in \textsc{SecSpace} or \textsc{EMMIE}, to support sharing is not readily available. Thus, a sharing channel needs to be initialized. The same threats of \textit{spoofing} and \textit{unauthorized access} from \textit{Personal Area Networks} such as ZigBee or Bluetooth arises.

\subsubsection*{Securing Sharing Channels} Likewise, similar techniques of \textit{out-of-band} channels can be used to achieve a secure channel initialization. \textsc{Looks Good To Me} (\textsc{LGTM}) is an authentication protocol for device-to-device sharing \cite{gaebel2016looks}. It is leveraged on the camera/s and wireless capabilities of existing AR HMDs. Specifically, it uses the combination of distance information through wireless localization and facial recognition information to cross-authenticate users. In other words, using the AR HMD that has a camera and wireless connectivity, users can simply look at each other to authenticate and initiate sharing. \textsc{HoloPair}, on the other hand, avoids the use of wireless localization, which may be unavailable (and inefficient) to most current devices, and instead utilizes exchange of \textit{visual cues} between users to confirm the shared secret \cite{sluganovic2017holopair}. Both uses the visual channel as an out-of-band channel.

\subsubsection{Remaining Challenges in Sharing and Interactions} The most apparent challenge are the varying use cases with which users interact or share. Depending on the context and/or situation, privacy and security concerns, as well as the degree of concern, can vary. For example, \textit{feed-through signalling} may be necessary in classroom scenarios to inform teachers when students enter and leave the classroom. However, there would also be occasions that it could be perceived too invasive or counter-intuitive, say, during military operations, i.e. negotiations in the field, and the like. Thus, there is a great deal of subjectivity to determine what is the most effective protection mechanism during sharing or interactions. And, perhaps, before everything else, we should ask first: ``Who or what are we protecting?''.

\subsection{Device Protection}\label{sec:device}

Given the capabilities of these MR devices, various privacy and security risks and concerns have been raised. Various data protection approaches have also been proposed in the previous subsections. To complement these approaches, the devices themselves have to be protected as well. There are two general aspects that needs protection in the device level: device access, and display protection.

\subsubsection{Threats to \textbf{Device Access}}\label{sssec:deviceAccess} The primary threats to device access are \textit{identity spoofing} and \textit{unauthorized access}.

\subsubsection*{Novel Authentication Strategies} Device access control ensures that authorized users are provided access while unauthorized ones are barred. Currently, password still remains as the most utilized method for authentication \cite{passwords}. To enhance protection, \textit{multi-factor} authentication (MFA) is now being adopted, which uses two or more independent methods for authentication. It usually involves the use of the traditional password method coupled with, say, a dynamic key that can be sent to the user via SMS, email, or voice call. The two-factor variant has been recommended as a security enhancement, particularly in on-line services like E-mail, cloud storage, e-commerce, banking, and social networks.

Aside from passwords are pin-based and pattern-based methods that are popular as mobile device authentication methods. A recent study \cite{george2017seamless} evaluated the usability and security of these established pin- and pattern-based authentication methods in virtual interfaces and showed comparable results in terms of execution time compared to the original non-virtual interface. Now, we take a look at other novel authentication methods that is leveraged on the existing and potential capabilities of MR devices.

\begin{enumerate}
	
	\item \textit{Gesture- and Active Physiological-based Authentication.} We look at the various possible gestures that can easily be captured by MR devices, specifically finger, hand, and head gestures. Mid-air finger and hand gestures have been shown to achieve an accuracy between 86-91\% (based on corresponsing accuracy from the equal error rate or EER) using a 3D camera-based motion controller, i.e. Leap Motion, over a test population of 200 users \cite{aslan2014mid}. A combination of head gestures and blinking gestures triggered by a series of images shown through the AR HMD have also been evaluated and promises an approximately 94\% of balanced accuracy rate in user identification over a population of 20 users \cite{rogers2015approach}. On the other hand, \textsc{Headbanger} uses head-movements triggered by an auditory cue (i.e. music) and achieved a True Acceptance Rate (TAR) of 95.7\% over a test population of 30 users \cite{li2016move}. Other possible gestures or active physiological signals, such as breathing \cite{chauhan2017breathprint}, are also potential methods.

	\item \textit{Passive Physiological-based Authentication.} Passive methods include physiological or biometric signals. \textit{Physiological-signal-based key agreement} or (PSKA) \cite{venkatasubramanian2010pska} used PPG features locked in a \textit{fuzzy-vault} for secure inter-sensor communications for body area networks or BAN. Despite existing MR devices not having PPG sensing capabilities, the PSKA method can be utilized for specific use cases when MR devices need to communicate with other devices in a BAN such as other wearables which can potentially be PPG sensing capable. On the other hand, \textsc{SkullConduct} \cite{schneegass2016skullconduct} uses the bone conduction capability of the Google Glass for user identification (with TAR of 97\%) and authentication (EER of 6.9\%). All these novel methods show promise on how latent gestures, physiological signals, and device capabilities can be leveraged for user identification and authentication.

	\item \textit{Multi-modal Biometric Authentication} combines two or more modes in a singular method instead of involving other methods or bands of communication is called multi-modal authentication. One multi-modal method combines facial, iris, and periocular information for user authentication and has an EER of 0.68\% \cite{raja2015multi}. \textsc{GazeTouchPass} combines gaze gestures and touch keys as a singular pass-key for smart phones to counter shoulder-surfing attacks on touch-based pass keys \cite{khamis2016gazetouchpass}. These types of authentication methods can readily be applied to MR devices that has gaze tracking and other near-eye sensors.
	
\end{enumerate}

\subsubsection{Threats to \textbf{Physical Interfaces}}\label{sssec:interface_protection} As discussed in \S\ref{sssec:userInputs} and \S\ref{sssec:secretDisplays}, MR interfaces are vulnerable from malicious inference which leads to \textit{disclosure} of \textit{input} activity, and/or \textit{output display information}. Currently available personal AR or MR see-through HMDs project or display content through lenses. The displayed content on the see-through lenses can leak and be observed externally. Visual capture devices, say, a camera, can be used to capture and extract information from the display leakage. External input interfaces suffer from the same inference and side-channel attacks such as shoulder-surfing.

\subsubsection*{Protection Approaches} 
\begin{figure}
	\centering
	\includegraphics[width=0.9\textwidth]{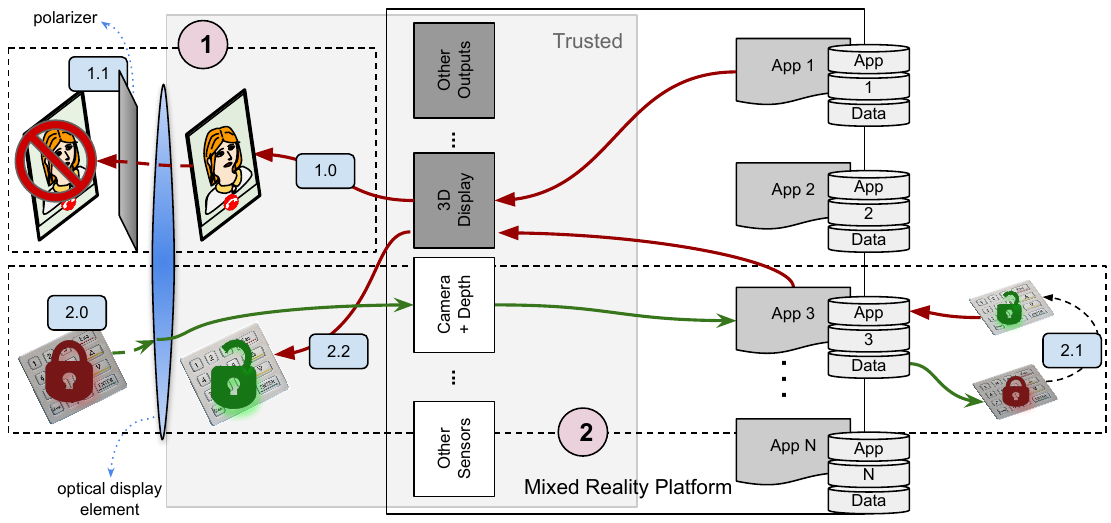}
	\caption{Sample interface and display protection strategies: 1) inserting a \textit{polarizer} to prevent or block display leakage; and 2) \textit{visual cryptography}, e.g. using secret augmentations (2.2) through decryption (2.1) of  encrypted public interfaces (2.0). All elements to the left of the \textit{optical display element} are considered \textit{vulnerable} to external inference or capture.}
	\label{fig:device_protection}
	\vspace{-3mm}
\end{figure}

There are optical and visual strategies that can be used to provide interface and activity \textit{confidentiality} and \textit{unobservability}. Figure \ref{fig:device_protection} shows example strategies of \textit{optical blocking} and \textit{visual cryptography}.

\begin{enumerate}
	\item \textit{Optical strategies} have been proposed, such as the use of \textit{polarization} on the outer layer (as in Figure \ref{fig:device_protection} labelled 1), use of \textit{narrowband illumination}, or a combination of the two to maximize display transmission while minimizing leakage  \cite{kohno2016display}. As of yet, this is the only work on MR display leakage protection using optical strategies.

	\subitem There are other capture protection strategies that have been tested on non-MR devices which allows objects to inherently or actively protect themselves. For example, the \textsc{TaPS} widgets use optical reflective properties of a scattering foil to only show content at a certain viewing angle, i.e. 90\degree \cite{mollers2011taps}.
	
	\subitem \textit{Active camouflaging techniques} have also been used, particularly on mobile phones, which allows the screen to blend with its surrounding just like a chameleon \cite{pearson2017chameleon}. Both \textsc{TaPS} widgets and the chameleon-inspired camouflaging are physically hiding sensitive objects or information from visual capture. The content-hiding methods discussed in \S\ref{sssec:secretDisplays} to hide outputs are also optical strategies.
	
	\item \textit{Visual cryptography} and {scrambling} techniques for display protection have also been discussed in \S\ref{sssec:secretDisplays}. The same can also be used for protecting sensitive input interfaces. \textsc{EyeDecrypt} \cite{forte2014eyedecrypt} uses \textit{visual cryptography technique} to protect input/output interfaces, say, ATM PIN pads, as shown in Figure \ref{fig:device_protection} labelled 2. The publicly viewable input interface is encrypted (step 2.0), and the secret key is kept or known by the user. The user uses an AR device to view the encrypted public interface and, through the secret key, is visually decrypted (step 2.1). As a result, only the user can see the actual input interface through the AR display (step 2.2). It utilizes \textit{out-of-band channels} to securely transmit the cryptographic keys between two parties (i.e. the client, through the ATM interface, and the bank). It also provides defence if the viewing device, i.e. the AR HMD, is untrusted by performing the visual decryption in a secure server rather than on the AR device itself.
	
	\subitem Another AR-based approach secretly scrambles keyboard keys to hide typing activity from external inference \cite{maiti2017preventing}. Only the user through the AR device can see the actual key arrangement of the keyboard. However, these techniques greatly suffer from visual alignment issues, i.e. aligning the physical display with the objects rendered through the augmented display. 

\end{enumerate}

\subsubsection{Remaining Challenges in Device Protection}
Despite the use-cases with visual cryptography using AR or MR displays, the usability of this technique is still confined to specific sensitive use cases due to the requirements of alignment. Also, this type of protection is only applicable to secrets that are pre-determined, specifically, information or activities that are known to be sensitive, such as password input or ATM PIN input. These techniques are helpful in providing security and privacy during such activities in shared or public space due to the secrecy provided by the near-eye displays which can perform the decryption and visual augmentation. Evidently, it only protects the output or displayed content of external displays but not the actual content which are displayed through the AR or MR device.

We have presented both defensive and offensive, as well as active and passive, strategies to device protection. Nonetheless, there are still numerous efforts on improving the input and output interfaces for these devices and it is opportune to consider in parallel the security and privacy implications of these new interfaces.

\subsection{Summary of Security and Privacy Approaches in Mixed Reality}\label{sec:summary} Finally, we present in Table \ref{tab:approaches_properties} an over-all comparison of these approaches based on which security and privacy properties they are addressing and to what extent, which can either be significant, partial or none.

\begin{table}
	\centering
	\caption{Summary of MR approaches that have been discussed, and which security and privacy properties are addressed by each approach and to what extent. The number of times a property is targeted is counted.}
	\label{tab:approaches_properties}
	\tiny
	\vspace{-2mm}
	\begin{tabular}{@{}cp{4.8cm}p{2.6mm}p{2.6mm}p{2.6mm}p{2.6mm}p{2.6mm}p{2.6mm}p{2.6mm}p{2.6mm}p{2.6mm}p{2.6mm}p{2.6mm}p{2.6mm}p{2.6mm}p{1mm}@{}}
		\toprule
		Section  & \textbf{Approach}     & \rotatebox{65}{Integrity}    & \rotatebox{65}{Non-Repudiation}    & \rotatebox{65}{Availability}    & \rotatebox{65}{Authorization}    &\rotatebox{65}{Authentication}    & \rotatebox{65}{Identification}   & \rotatebox{65}{Confidentiality}    & \rotatebox{65}{Anonymity}   & \rotatebox{65}{Unlinkability}    & \rotatebox{65}{Undetectability}   &\rotatebox{65}{Deniability} &\rotatebox{65}{Awareness} &\rotatebox{65}{Compliance}  \\ \midrule
		
		& {{Input Protection Approaches}} &    0     &   3     &   0     &   12     &    0    &   0     &  14 &   13     &   0     &  14   &0 &  12 & 14     \\ \midrule
		\ref{sssec:sanitization}	&  {\tiny\textsc{Darkly} \cite{jana2013scanner}}$\ddagger$   &         & $\checkmark$      &   & {\tiny$\checkmark\checkmark$}        &   &        & $\checkmark$      & $\checkmark$      &        & $\checkmark$      &     & {\tiny$\checkmark\checkmark$}      & $\checkmark$    \\
		\ref{sssec:sanitization}	&  {\tiny\textit{Context-based sanitization} \cite{zarepour2016context}}$\ddagger$  &           &        &      & $\checkmark$      &   &        & $\checkmark$      & $\checkmark$      &        & $\checkmark$      &      & {\tiny$\checkmark\checkmark$}       & $\checkmark$   \\
		\ref{sssec:sanitization}	&  {\tiny\textsc{PlaceAvoider} \cite{templeman2014placeavoider}}$\dagger$ &           &        &      & $\checkmark$      &   &        & $\checkmark$      & $\checkmark$      &        & $\checkmark$      &      & {\tiny$\checkmark\checkmark$}      & $\checkmark$   \\
		\ref{sssec:sanitization}	&  {\tiny\textit{3D humanoids replace humans} \cite{szczuko2014augmented}}$\ddagger$  &         &        &        &        &      &        & $\checkmark$      & $\checkmark$ &        & $\checkmark$      &        & $\checkmark$      & $\checkmark$      \\
		\ref{sssec:sanitization}	& {\tiny\textsc{OpenFace/RTFace} \cite{wang2017scalable}}$\dagger$   &         &        &        &        &      &        & $\checkmark$      & $\checkmark$ &        & $\checkmark$      &        & $\checkmark$      & $\checkmark$      \\
		\ref{sssec:sanitization}	&  {\tiny\textit{Capture-resistant spaces} \cite{truong2005preventing}}$^{\circ}$   &         &        &        &     $\checkmark$  &  &        & $\checkmark$      & $\checkmark$ &        & $\checkmark$      &        &        & $\checkmark$      \\
		\ref{sssec:sanitization}	&   {\tiny\textit{See-through vision} \cite{hayashi2010installation}}$\ddagger$  &         &        &        & $\checkmark$      &      &        & $\checkmark$      & $\checkmark$ &        & $\checkmark$      &        & $\checkmark$      & $\checkmark$      \\
		\ref{sssec:sanitization}	&   {\tiny\textit{World-driven access control} \cite{roesner2014world}}$\ddagger$  &         &        &        & $\checkmark$      &      &        & $\checkmark$      & $\checkmark$ &        & $\checkmark$      &        & $\checkmark$      & $\checkmark$      \\
		\ref{sssec:sanitization} 	&  {\tiny\textsc{I-pic} \cite{aditya2016pic}}$\dagger$   &         &        &        & $\checkmark$      &      &        & $\checkmark$      & $\checkmark$ &        & $\checkmark$      &        & $\checkmark$      & $\checkmark$      \\
		\ref{sssec:sanitization}  	&   {\tiny\textsc{PrivacyCamera} \cite{li2016privacycamera}}$\dagger$   &         &        &        & $\checkmark$      &      &        & $\checkmark$      & $\checkmark$ &        & $\checkmark$      &        & $\checkmark$      & $\checkmark$      \\
		\ref{sssec:sanitization}         &  {\tiny\textsc{Cardea} \cite{shu2016cardea}}$\dagger$    &         &        &        & $\checkmark$      &      &        & $\checkmark$      & $\checkmark$ &        & $\checkmark$      &        & $\checkmark$      & $\checkmark$      \\
		\ref{sssec:sanitization}       &  {\tiny\textsc{MarkIt} \cite{raval2014markit,raval2016you}}$\dagger$  &         &        &        & $\checkmark$      &      &        & $\checkmark$      & $\checkmark$ &        & $\checkmark$      &        & $\checkmark$      & $\checkmark$      \\
		\ref{sssec:userInputs}   &  {\tiny\textsc{PrePose} \cite{figueiredo2016prepose}}$\ddagger$   &        & $\checkmark$      &        & {\tiny$\checkmark\checkmark$}      &        &        & {\tiny$\checkmark\checkmark$}      &        &        & $\checkmark$      &     &        & $\checkmark$    \\
		\ref{sssec:userInputs}  &   {\tiny\textsc{Recognizers} \cite{jana2013enabling}}$\ddagger$  &     & $\checkmark$      &   & {\tiny$\checkmark\checkmark$}      &   &        & {\tiny$\checkmark\checkmark$}      & $\checkmark$      &        & $\checkmark$      &     & $\checkmark$      & $\checkmark$    \\ \midrule
		
		& {{Data Protection Approaches}} &    8     &   1     &   5     &   5     &    0    &   0     &  11 &   5    &   2     &  8   & 7&  2 & 2    \\ \midrule
		 \ref{ssec:data_collection}      & {\tiny\textit{Randomized Response }\cite{erlingsson2014rappor}}$^{\circ}$    &    &       &   &      &     &   &   {\tiny$\checkmark\checkmark$}   &    $\checkmark$    &    $\checkmark$   &  $\checkmark$   &     {\tiny$\checkmark\checkmark$}     & $\checkmark$      & $\checkmark$      \\
		\ref{ssec:data_collection}      &  {\tiny\textsc{SemaDroid}\cite{xu2015semadroid}}$^{\circ}$    &    & $\checkmark$      &   & {\tiny$\checkmark\checkmark$}     &     &   & $\checkmark$      &            &       &   $\checkmark$  &         & $\checkmark$      & $\checkmark$      \\
		\ref{ssec:data_processing}      &  {\tiny\textsc{HE-sift} \cite{jiang2017secure}}$^{\circ}$   & $\checkmark$       &        & $\checkmark$      &        &    &    & {\tiny$\checkmark\checkmark$}      &       &        & $\checkmark$      & $\checkmark$      &        &            \\
		\ref{ssec:data_processing}      &  {\tiny\textit{Leveled} \textsc{HE-sift}\cite{zhang2014cloud}}$^{\circ}$    & $\checkmark$       &        & $\checkmark$      &        &    &    & {\tiny$\checkmark\checkmark$}      &      &        & $\checkmark$      & $\checkmark$      &        &            \\
		\ref{ssec:data_processing}      &  {\tiny\textsc{CryptoImg} \cite{ziad2016cryptoimg}}$^{\circ}$   & $\checkmark$       &        &        &        &   &     & {\tiny$\checkmark\checkmark$}      &  &        & $\checkmark$      & $\checkmark$      &        &         \\
		\ref{ssec:data_processing}      &  {\tiny\textsc{SecSIFT} \cite{qin2014private,qin2014towards,qin2014privacy}}$^{\circ}$      & $\checkmark$       &        &        &        &    &    & {\tiny$\checkmark\checkmark$}      &    &        & $\checkmark$      & $\checkmark$      &        &         \\
		\ref{ssec:data_processing}      &  {\tiny\textsc{P3} \cite{ra2013p3}}$^{\circ}$      & $\checkmark$       &        &        &        &   &     & {\tiny$\checkmark\checkmark$}      &   &        & $\checkmark$      & $\checkmark$      &        &           \\
		\ref{ssec:data_processing}      &  {\tiny\textit{Cloth Try-on} \cite{sekhavat2017privacy}}$\ddagger$  &      &        &      & $\checkmark$      &        &        & {\tiny$\checkmark\checkmark$}       & $\checkmark$      & $\checkmark$ & {\tiny$\checkmark\checkmark$}   & $\checkmark$      &   &   \\
		\ref{ssec:data_storage}      &  {\tiny\textsc{PDV} \cite{mun2010pdv}}$^{\circ}$    & {\tiny$\checkmark\checkmark$}       &        & $\checkmark$      & {\tiny$\checkmark\checkmark$}      &      &      & $\checkmark$ & $\checkmark$      &        &        &      &        &      \\
		\ref{ssec:data_storage}      &  {\tiny\textsc{OpenPDS} \cite{deMontjoye2014openpds}}$^{\circ}$   & {\tiny$\checkmark\checkmark$}       &        & $\checkmark$      & {\tiny$\checkmark\checkmark$}      &      &      & {\tiny$\checkmark\checkmark$}& $\checkmark$      &        &        &      &        &      \\
		\ref{ssec:data_storage}      &  {\tiny\textsc{DataBox}\cite{crabtree2016databox}}$^{\circ}$   & {\tiny$\checkmark\checkmark$}       &        & $\checkmark$      & {\tiny$\checkmark\checkmark$}      &      &      & {\tiny$\checkmark\checkmark$} & $\checkmark$      &        &        &      &        &      \\   \midrule
		
		& {Output Protection Approaches} &    1     &   1     &   4     &   4     &    0    &   0     &  8 &   2     &   0     &  6   &0 &  2 & 2     \\ \midrule
		\ref{sssec:safeOutputs}   &  {\tiny\textsc{Arya} \cite{lebeck2016safely, lebeck2017securing}}$\ddagger$    & $\checkmark$  & $\checkmark$       & $\checkmark$      & {\tiny$\checkmark\checkmark$}         &        &        & $\checkmark$      &        &        &        &    & $\checkmark$      & {\tiny$\checkmark\checkmark$}      \\
		\ref{sssec:secureRendering}   &   {\tiny\textsc{SurroundWeb} \cite{vilk2014least, vilk2015surroundweb}}$\ddagger$    &       &         &        & $\checkmark$    &        &        & $\checkmark$      & $\checkmark$      &        & {\tiny$\checkmark\checkmark$}       &     & $\checkmark$      & {\tiny$\checkmark\checkmark$}    \\ 
		\ref{sssec:secretDisplays}      &   {\tiny\textsc{EyeGuide} \cite{eaddy2004my}}$\ddagger$   &      &        &        &{\tiny$\checkmark\checkmark$}   &        &   & {\tiny$\checkmark\checkmark$}   &        &        & {\tiny$\checkmark\checkmark$}      &    &        &       \\
		\ref{sssec:secretDisplays}      &   {\tiny\textsc{VR Codes} \cite{woo2012vrcodes}}$\ddagger$      &       &        & $\checkmark$      &      &        &      & $\checkmark$    &        &        & {\tiny$\checkmark\checkmark$}      &     &        &      \\
		\ref{sssec:secretDisplays}     &   {\tiny\textit{Psycho-visual modulation} \cite{lin2017video}}$\ddagger$  &       &        & $\checkmark$      &      &        &      & $\checkmark$    &        &        & {\tiny$\checkmark\checkmark$}      &     &        &      \\
		\ref{sssec:secretDisplays}    &  {\tiny\textit{Visual cryptography using AR} \cite{lantz2015visual}}$\ddagger$   &        &        &        &     &        &        & {\tiny$\checkmark\checkmark$}   &        &        & {\tiny$\checkmark\checkmark$}      &    &        &       \\
		\ref{sssec:secretDisplays}   &   {\tiny\textit{Secret messages using AR} \cite{andrabi2015usability}}$\ddagger$  &        &        &        &     &        &        & {\tiny$\checkmark\checkmark$}   &        &        & {\tiny$\checkmark\checkmark$}      &    &        &       \\
		\ref{sssec:secretDisplays}   &  {\tiny\textit{Mobile visual cryptography} \cite{fang2010securing}}$\ddagger$  &        &        & $\checkmark$      & {\tiny$\checkmark\checkmark$}     &        &        & $\checkmark$ & $\checkmark$ &        &        &     &        &      \\ \midrule
%
%
%
%
		& {Interaction Protection Approaches}  &    2     &   4     &   4     &   6     &    2    &   4     &  9 &   1     &   0     &  5   &0 &  9 & 7     \\ \midrule
		\ref{sssec:collabInteractions}      &   {\tiny\textsc{Emmie} \cite{butz1998vampire, butz1999emmie}}$\ddagger$  &         &        &        &     $\checkmark$   &        &        &  $\checkmark$ &        &        &  $\checkmark$    &     &   $\checkmark$     &   {\tiny$\checkmark\checkmark$}   \\
		\ref{sssec:collabInteractions}     &  {\tiny\textsc{Kinected Conference}\cite{devincenzi2011kinected}}$\ddagger$  &        &        &        &      &        &        &  $\checkmark$ &    $\checkmark$    &        &    $\checkmark$  &      &   $\checkmark$     &   {\tiny$\checkmark\checkmark$}  \\
		\ref{sssec:collabInteractions}    &  {\tiny\textsc{SecSpace} \cite{reilly2014secspace}}$\ddagger$  &      &    $\checkmark$    &     $\checkmark$   &      &        &    $\checkmark$    &  $\checkmark$ &        &        &      &     &      {\tiny$\checkmark\checkmark$}  &    {\tiny$\checkmark\checkmark$} \\
		\ref{sssec:collabInteractions} &  {\tiny\textit{Candid Interactions} \cite{ens2015candid}}$^{\circ}$  &      &    $\checkmark$    &     $\checkmark$   &      &        &    $\checkmark$    &  $\checkmark$ &        &        &      &     &      $\checkmark$  &    $\checkmark$ \\
		\ref{sssec:collabInteractions}    &   {\tiny\textsc{Pip} \cite{szalavari1997personal, szalavari1998gaming}}$\ddagger$ &         &        &        &     $\checkmark$   &        &        &  $\checkmark$ &        &        &  $\checkmark$    &     &   $\checkmark$     &   $\checkmark$  \\
		\ref{sssec:collabInteractions}    &   {\tiny\textsc{TouchSpace} \cite{cheok2002touch}}$\ddagger$ &         &        &        &     $\checkmark$   &        &        &  $\checkmark$ &        &        &  $\checkmark$    &     &   $\checkmark$     &   $\checkmark$   \\
		\ref{sssec:collabInteractions}    &  {\tiny\textsc{BragFish} \cite{xu2008bragfish}}$\ddagger$   &         &        &        &     $\checkmark$   &        &        &  $\checkmark$ &        &        &  $\checkmark$    &     &   $\checkmark$     &   $\checkmark$   \\
		\ref{sssec:sharing}    &  {\tiny\textit{LooksGoodToMe}\cite{gaebel2016looks}}$\dagger$   &   $\checkmark$      &     {\tiny$\checkmark\checkmark$}   &      $\checkmark$  &     $\checkmark$   &    {\tiny$\checkmark\checkmark$}    &     {\tiny$\checkmark\checkmark$}   &  {\tiny$\checkmark\checkmark$} &        &   &        &   &    $\checkmark$    &    \\ 
		\ref{sssec:sharing}    &  {\tiny\textsc{HoloPair} \cite{sluganovic2017holopair}}$\dagger$   &   $\checkmark$      &     {\tiny$\checkmark\checkmark$}   &      $\checkmark$  &     $\checkmark$   &    {\tiny$\checkmark\checkmark$}    &     {\tiny$\checkmark\checkmark$}   &  {\tiny$\checkmark\checkmark$} &        &   &        &   &    $\checkmark$    &    \\  \midrule
		
		& {Device Protection Approaches}  &    2     &   0     &   0     &   11     &    8    &   9     &  13 &   0     &   0     &  3   &0 &  0 & 0     \\ \midrule
		\ref{sssec:deviceAccess}   &  {\tiny\textit{Seamless \& secure VR} \cite{george2017seamless}}$\ddagger$  &       &        &        &        &    {\tiny$\checkmark\checkmark$}    &   $\checkmark$     &   $\checkmark$    &        &        &        &    &        &       \\
		\ref{sssec:deviceAccess}       &  {\tiny\textit{Mid-air authentication gestures} \cite{aslan2014mid}}$\dagger$  &        &   &         &    $\checkmark$    &   {\tiny$\checkmark\checkmark$}    &     $\checkmark$   &     $\checkmark$  &        &        &        &     &        &      \\
		\ref{sssec:deviceAccess}       &  {\tiny\textit{Head and blinking gestures} \cite{rogers2015approach}}$\dagger$  &        &   &         &    $\checkmark$    &   {\tiny$\checkmark\checkmark$}    &     $\checkmark$   &     $\checkmark$  &        &        &        &     &        &      \\
		\ref{sssec:deviceAccess}    &  {\tiny\textsc{HeadBanger} \cite{li2016move}}$\ddagger$  &        &   &         &    $\checkmark$    &   {\tiny$\checkmark\checkmark$}    &     $\checkmark$   &     $\checkmark$  &        &        &        &     &        &      \\
		\ref{sssec:deviceAccess}  &  {\tiny\textsc{Pska} \cite{venkatasubramanian2010pska}}$^{\circ}$  &        &   &         &    $\checkmark$    &   {\tiny$\checkmark\checkmark$}    &     $\checkmark$   &     $\checkmark$  &        &        &        &     &        &      \\
		\ref{sssec:deviceAccess}      & {\tiny\textsc{SkullConduct} \cite{schneegass2016skullconduct}}$\dagger$ &        &   &         &    $\checkmark$    &   {\tiny$\checkmark\checkmark$}    &     $\checkmark$   &     $\checkmark$  &        &        &        &     &        &      \\
		\ref{sssec:deviceAccess}    &  {\tiny\textit{Facial multi-modal authentication} \cite{raja2015multi}}$\dagger$  &        &   &         &    $\checkmark$    &   {\tiny$\checkmark\checkmark$}    &     $\checkmark$   &     $\checkmark$  &        &        &        &     &        &      \\
		\ref{sssec:deviceAccess}      &  {\tiny\textsc{GazeTouchPass} \cite{khamis2016gazetouchpass}}$\dagger$   &        &   &         &    $\checkmark$    &   {\tiny$\checkmark\checkmark$}    &     $\checkmark$   &     $\checkmark$  &        &        &        &     &        &      \\
		\ref{sssec:interface_protection}     &   {\tiny\textit{Polarization} \cite{kohno2016display}}$\ddagger$   &        &        &        &           &        &      &   {\tiny$\checkmark\checkmark$}     &        &        &  {\tiny$\checkmark\checkmark$}     &     &        &      \\
		\ref{sssec:interface_protection}    &  {\tiny\textsc{Taps} Widget \cite{mollers2011taps}}$^{\circ}$   &         &        &        &   $\checkmark$     &        &        &     $\checkmark$   &        &      &      $\checkmark$  &  &        &           \\
		\ref{sssec:interface_protection}   &  {\tiny\textit{Chameleon-like} \cite{pearson2017chameleon}}$^{\circ}$   &         &        &        &      $\checkmark$  &        &        &      $\checkmark$  &        &      &      $\checkmark$  &        &        &      \\ 
		\ref{sssec:interface_protection}   &   {\tiny\textsc{EyeDecrypt} \cite{forte2014eyedecrypt}}$\ddagger$   &$\checkmark$      &        &        & {\tiny$\checkmark\checkmark$}      &   & $\checkmark$ & {\tiny$\checkmark\checkmark$}  &   &   &      &    &        &     \\
		\ref{sssec:interface_protection}    &   {\tiny\textit{Preventing keystroke inference} \cite{maiti2017preventing}}$\ddagger$  & $\checkmark$     &        &        & {\tiny$\checkmark\checkmark$}     &        &      & {$\checkmark$}  &        &        &        &     &        &      \\
		\midrule \midrule
		\multicolumn{2}{r}{Total Count}  &    13     &   9   &   13     &   38     &    10    &   13     &  55 &   21    &   2     &  36  &7 &  25 & 25    \\
		\bottomrule
	\end{tabular}
	\tiny
	\begin{tabular}{p{0.95\textwidth}}
				The extent of each approach was either {\tiny$\checkmark\checkmark$} significantly addressing, {\small$\checkmark$} partially addressing, or not (to minimally) addressing the security and privacy properties.
				The approaches have been applied to either an $\ddagger$ MR context, a $\dagger$ proto-MR context, or a $^{\circ}$ non-MR context. 
	\end{tabular}
\end{table}

\subsubsection*{Generalizations and gaps} Unsurprisingly, all approaches are targeting \textit{confidentiality} or preventing \textit{information disclosure} but others are much better at achieving it than most. However, the approach may only be providing significant protection on their specific space or target entity. For example, among the input protection approaches, only two provide significant confidentiality: \textsc{PrePose} only provides gesture events to requiring applications while raw input capture feed is kept hidden; and \textsc{Recognizers-}approach only provides the necessary recognizer data to requiring applications while keeping raw RGB and depth information hidden. But they do not provide any further protection outside their target scope.

Aside from confidentiality, most of the input protection approaches are providing \textit{authorization} or \textit{access control}, \textit{undetectability}, \textit{policy compliance}, \textit{anonymity}, and \textit{content awareness}. It is to no surprise as well because most of the approaches are implementing \textit{media sanitization} (\S\ref{sssec:sanitization}) and/or \textit{access control} (\S\ref{sssec:userInputs}) to provide \textit{anonymity} and \textit{undetectability} of users inputs, and bystanders. However, none of the approaches are targeting \textit{unlinkability} despite the provided anonymity and undetectability. This is due to the fact that these two properties are only provided through the access control or abstractions by most approaches. Once applications have been provided access to the media, resource, or their corresponding abstractions, anonymity and undetectability no longer hold and, thus, unlinkability as well. Therefore, further protection is necessary once data flows beyond the input interfaces.

To provide a protection that targets a combination of anonymity, undetectability, and unlinkability, most strategies use techniques that ensure statistical privacy guarantees such as \textit{differential privacy} and k\textit{-anonymity}. Among the data protection approaches listed in the summary, only \textsc{Rappor} and the \textit{virtual cloth try-on} provides a combined protection on those three properties: \textsc{Rappor} uses \textit{randomised response} to provide differential privacy during data collection, while the privacy-preserving virtual cloth try-on uses \textit{secure two-party computation}. Other data protection approaches, specifically, \textit{protected data processing} (\S\ref{ssec:data_processing}) primarily provides data \textit{integrity}, \textit{undetectability}, and \textit{plausible deniability} through cryptographic approaches, while \textit{protected data storage} (\S\ref{ssec:data_storage}) primarily provides \textit{integrity}, \textit{availability}, and \textit{anonymity} through \textit{authorized access} to user data.

Output protection approaches are primarily targeting \textit{undetectability} of outputs through visual hiding or cryptography (\S\ref{sssec:secretDisplays}), while \textit{abstraction} strategies in output (\S\ref{sssec:safeOutputs} and \ref{sssec:secureRendering}), just like in input abstraction, provide \textit{undetectability} through enforced \textit{policy compliance} and \textit{authorized access} to output or rendering resources as well as output data \textit{awareness}. Output policies ensure the \textit{availability}, and \textit{reliability} of outputs which, then, leads to user \textit{safety}.

\textit{Awareness} and \textit{policy \& consent compliance} are two other paramount properties targeted by the interaction protection approaches. Their primary goal is to deliver fair interactions among users despite of the `boundaries' that exist in shared spaces (\S\ref{sssec:collabInteractions}) either through \textit{user-mediated} information release or \textit{feed-forwarding}. Strategies on sharing initialization in MR (\S\ref{sssec:sharing}) primarily focused on security properties of \textit{non-repudiation}, \textit{authentication}, and \textit{identification} as well as \textit{integrity}, \textit{availability}, and \textit{authorization}.

Lastly, device protection approaches primarily target \textit{authorization}, \textit{identification}, and \textit{confidentiality} through\textit{authentication} strategies that capitalizes on MR capabilities. In addition, \textit{visual} protection strategies focuses on providing physical \textit{undetectability} and prevents \textit{unauthorized inference} from MR displays and devices.

The categorization, to some extent, has localised the targeted properties by the approaches. Together, the first three categories roughly target most properties but are slightly \textit{privacy} leaning especially for the input protection approaches. On the other hand, the last two categories were more \textit{security} leaning especially for the device protection strategies. The concentration of check marks ({\tiny$\checkmark\checkmark$}, {\small$\checkmark$}) highlights these preferred properties by the categories.

\begin{figure}
	\centering
	\includegraphics[width=\textwidth]{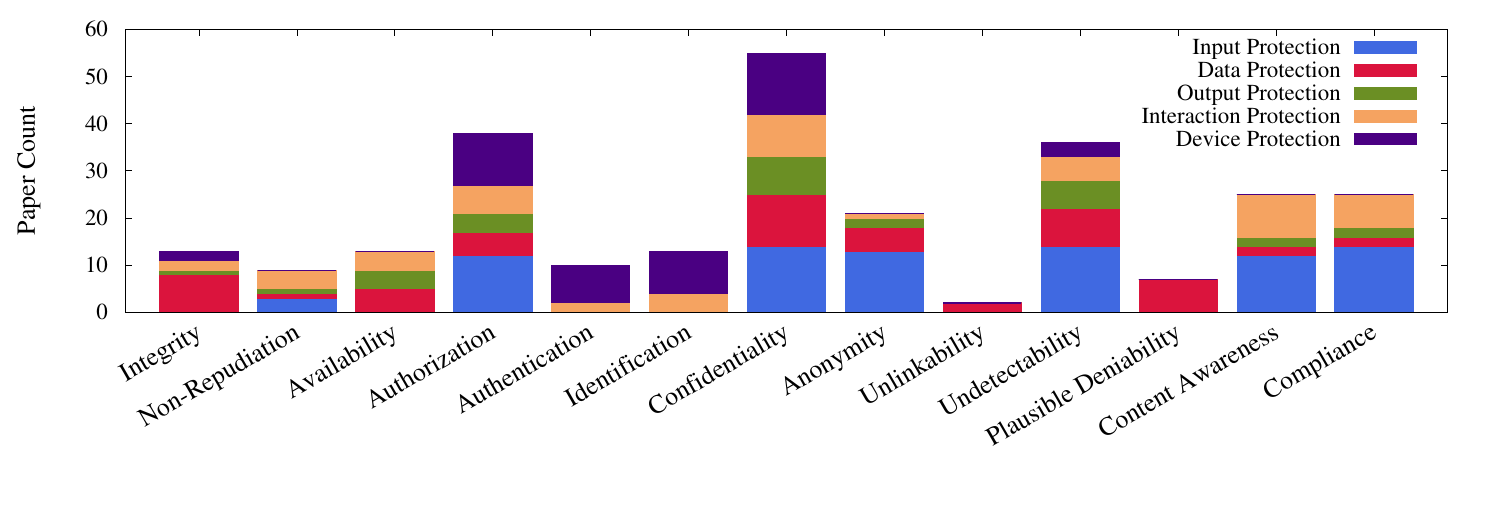}
	\vspace{-10mm}
	\caption{Distribution of categories among the thirteen properties.}
	\label{fig:paperDistribution}
	\vspace{-3mm}
\end{figure}

Figure \ref{fig:paperDistribution} visually shows the distribution of the approaches among the thirteen properties sorted according to their category. Excluding confidentiality, the most targeted properties are \textit{authorization} and \textit{undetectability \& unobservability}. The next most targeted are \textit{content awareness} and \textit{policy \& consent compliance}. 
Consequently, it is evident with this visualized graph that protection approaches still primarily lack provisions for \textit{unlinkability} and \textit{plausible deniability}. It is of no surprise that data protection approaches (which are mostly generic, non-MR targeted) are the ones that primarily target these two properties. Thus, there is a great opportunity to explore, investigate, design, and develop strategies for MR that targets them on top of the evidently targeted properties of existing approaches.

\section{Open challenges}\label{sec:challenges}
The previous section has elaborated on the different security and privacy approaches in the research literature, and highlighted the remaining challenges for each category. Now, we present the high-level open challenges brought about by existing and upcoming MR devices, and our potential future with MR.

\paragraph{\textbf{Security and Privacy of Existing Devices and Platforms}}Most of the security and privacy work discussed in \S\ref{sec:s&p} have only been applied on predecessor or prototype platforms of MR. Now, we take a broader look on the security and privacy aspects of existing devices and platforms. 
Most of the current tool kits, platforms, and devices have not considered the security and privacy risks that were highlighted in this survey. There was an effort to pinpoint risks and identify long-term protection for some platforms, specifically Wikitude, and Layar \cite{McPherson2015ARBrowsers} but no actual nor systematic evaluation was done on both the risks and the proposed protection.

A \textit{systematic security and privacy analysis} of MR applications, devices, and platforms has to be performed in order to identify other potential and latent risks. For example, Hololens' 3D scanning capability should be investigated whether its resolution can potentially be used to detect heartbeats or other physiological signals of bystanders which are considered to be sensitive information. Then, we can evaluate these systems against further security and privacy requirements that address beyond the known threats and exploits. Upon consideration of these requirements, we can, then, come up with protection mechanisms targeted for these devices and platforms.

\paragraph{\textbf{Regulation of Mixed Reality applications}} Moreover, the current MR devices and platforms now provides a peek on what the future devices will offer. Given the capabilities of these devices, both in sensing and rendering, there are numerous concerns on security and privacy that have been raised and needs to be addressed. In fact, real vulnerabilities have already been revealed particularly on the numerous available MR applications 
For example, Pokemon Go's network messages between the application and their servers on its early releases were vulnerable to man-in-the-middle attacks \cite{colceriu2016catching} which allows hackers to get precise location information of users. In addition to these were the numerous malicious versions of the application \cite{burgess2016pokemongo,frink2016pokemongo} that has been downloaded and used by many users. Thus, there is a need for \textit{regulation}, whether applied locally on the devices or platforms, or in a larger scale through a much stringent \textit{reputation}- and \textit{trust}-based \textit{ecosystem} of applications.

\paragraph{\textbf{Native Support}} Furthermore, the entire MR pipeline processing (of detection, transformation, and rendering as shown in Figure \ref{fig:MRpipeline}) is performed by the applications themselves and in a very monolithic manner. They would need direct access to sensing interfaces to perform detection, and to output interfaces for rendering. Once provided permission, AR applications, inevitably, have indefinite access to these resources and their data. Native, operating system-level support for detection and rendering can provide an access control framework to ensure security and privacy of user data. OS-level abstractions have been proposed \cite{dantoni2013} to ``expose only the data required to [third-party] applications''. Apple's ARkit is one of those moves to provide native support to MR application development on their platforms, but it's granularity of access control is still coarse and similar risks are still present.

\paragraph{\textbf{Targeted Data Protection}} As highlighted in \S\ref{sec:summary}, there is a lack of MR-targeted data protection approaches except for one specific use case on virtual cloth try-ons. Due to the highly rich and dynamic data that is associated to MR, generic data protection approaches may not be readily applicable and translatable to MR systems. Thus, there is a huge necessity to design and evaluate data protection approaches (e.g. encryption, privacy-preserving algorithms, and personal data stores) targeted for MR data.


\paragraph{\textbf{User welfare, society, policy, and ethics}}
Possible social consequences \cite{feiner1999wearableAR} may arise from the user's \textit{appearance} while wearing the device, and, further, implications on bystander privacy which have already been brought up several times. Existing (universal, ethical, and legal) concepts on privacy and policy may need to be revisited to catch up with MR becoming mainstream. As these technologies will be delivering information overlaid on the physical space, the correctness, safety, and legality of this information has to be ensured \cite{roesner2014policy}. This is now more evident with demonstrations such as the \textit{Google Duplex}, and their \textit{Visual Positioning System}\footnote{Both were demonstrated during Google I/O 2018}.\

Other user welfare challenges include the reduction of physical burden from wearing the device; eye discomfort from occlusion problems, incorrect focus, or output jitter; and the \textit{cognitive disconnect} from the loss of tactile feedback when ``interacting'' with the virtual objects.

\paragraph{\textbf{Profitability and Interoperability}}
Ultimately, before MR becomes widely adopted, interoperability and profitability are additional aspects that needs to be considered. Manufacturers and developers has to make sure the profitability of these devices once delivered to the market \cite{kroeker2010mainstreaming}. Subsequently, ensuring the interoperability of these devices with existing technologies is key to user adoption of AR and MR devices. There are already demonstrations of how web browsing can be interoperable between AR and non-AR platforms \cite{lee2009interoperable}. As future and upcoming MR services will be web-based or web-assisted, this can serve as a basis.

\paragraph{\textbf{Networking Challenges}}
Due to the expected increase of MR applications, services, platforms, and devices, network demand is also expected to increase for these services to be delivered to the users at an acceptable latency. And it is not only confined to delivery or transmission of MR data, but also to processing (i.e. data transformation) and rendering. Current data rates that are provided by existing communication networks may not be ready for such demand \cite{braud2017network}. Thus, it is an imperative that current and upcoming network design and infrastructure has to be rethought and remains a challenge.

\paragraph{\textbf{Smart Future}}
Overall, the same with all the other technologies that try to push through the barriers, challenges on processing, data storage, communication, energy management, and battery still remains at the core. Despite all these, a \textit{smart world} future is still in sight where smart objects integrate to form smart spaces that, then, form smart \textit{hyperspaces} \cite{ma2005smartworld}. This \textit{smart world} will be composed of `things' with enough intelligence to be \textit{self-aware} and manage its own processing, security, and energy. MR has been shown to enable a \textit{cross-reality} space \cite{lifton2009metaphor} -- freely crossing from physical to synthetic and vice versa -- that can allow these self-aware devices to \textit{understand} the real-world while managing all information \textit{virtually}. Lastly, we reiterate that existing networks have to be analysed and evaluated if it can support this future and should be redesigned if necessary.

\section{Conclusion}\label{sec:conclusion}
This is the first survey to take in the endeavour of collecting, categorizing, and reviewing various security and privacy approaches in MR. We have raised various known and latent security and privacy risks associated with the functionalities of MR and gathered a comprehensive collection of security and privacy approaches on MR and related technologies. We have identified five aspects of protection namely input, data access, output, interactivity, and device integrity, and categorized the approaches according to these five aspects. Furthermore, we identify which security and privacy properties are targeted by the approaches using a list of thirteen known properties. We, then, used these properties to present a high-level description of the approaches and use it to compare them. Among the properties, aside from \textit{confidentiality}, \textit{authorization} and \textit{undetectability} were the most targeted, while there is a lack in the provision of other properties such as \textit{unlinkability} and \textit{plausible deniability}. Therefore, it is opportune to design, investigate, and implement security and privacy mechanisms that can be integrated with existing and upcoming MR systems, while their utilization and adoption are still not widespread.

\begin{screenonly}
\appendix

\setcounter{table}{0}
\renewcommand{\thetable}{\textbf{A\arabic{table}}}

\section{Mixed Reality Hardware}\label{asec:hardware}

Table \ref{tab:devices} lists some commercially available HMDs. Most of the devices listed are \textit{partially see-through devices} such as the Recon Jet\footnote{Recon Jet, (Accessed 4 Aug 2017), https://www.reconinstruments.com/products/jet}, and the Vuzix M100/M300\footnote{Products, (Accessed 4 Aug 2017), https://www.vuzix.com/Products} smart glasses. These smart glasses sprung from the wearable space and are, possibly, the most popularly known and widely used types of HMDs. They use a small enough video display positioned near one eye (usually on the right) so as not to entirely block the user's FOV. This is in comparison to \textit{video see-through} HMDS which employ a video camera usually front-facing, as in the HTC Vive\footnote{Buy VIVE Hardware, (Accessed 4 Aug 2017), https://www.vive.com/au/product/}, to capture the environment, then, feed the video to a display propped in front of the user's field of view (FOV).

\begin{table}
	\centering
	\scriptsize
	\caption{Commercially available MR HMDs}
	\label{tab:devices}
	\setlength{\extrarowheight}{.5em}
	\begin{tabular}{@{}p{1.5cm}p{2.5cm}p{1.7cm}p{1.3cm}p{1cm}p{1.7cm}p{2cm}@{}}
		\toprule
		\textbf{Device}                     & \textbf{Sensors}                                                                                                                                & \textbf{Controls or UI}                                                                                                 & \textbf{Processor}                                                 & \textbf{OS}          & \textbf{Outputs}                                                                            & \textbf{Connectivity}                                                  \\ \midrule
		Google Glass 1.0 Explorer Edition   & Camera (5MP still/ 720p video), Gyroscope, accelerometer, compass, light and proximity sensor, microphone                                       & Touchpad, remote control via app (iOS, Android)                                                                                     & TI OMAP 4430 1.2GHz dual core (ARMv7)                  & Android 4.4                                       & 640x360 (LCoS, prism), Bone coduction audio transducer                                      & WiFi 802.11 b/g, Bluetooth, microUSB                                   \\ 
		Google Glass 2.0 Enterprise Edition & Camera (5MP still/ 720p video), Gyroscope, accelerometer, compass, light and proximity sensor, microphone                                       & Touchpad, remote control via app (iOS, Android)                                                                                     & Intel Atom (32-bit)                                                  & Android 4.4                                       & 640x360 (LCoS, prism), Audio speaker                                                        & WiFi 802.11 dual-band a/b/g/n/ac, Bluetooth LE, microUSB \\ 
		ReconJet Pro                        & Camera, Gyroscope, accelerometer, compass, pressure, IR                                                                                         & 2 buttons, 4-axis optical touchpad                                                                                                  & 1GHz dual-core ARM Cortex-A9                                         & ReconOS 4.4 (Android-based)                    & 16:9 WQVGA, Audio speaker                                                                   & WiFi 802.11 b/g, Bluetooth, GPS, microUSB                              \\ 
		Vuzix M300                          & Camera (13MP still/1080p video), proximity (inward and outward), gyroscope, accelerometer, compass, 3-DoF head tracking, dual microphones       & 4 buttons, 2-axis touchpad, remote control via app (iOS, Android), voice                                                            & Dual-core Intel Atom                                                 & Android 6.0                       & 16:9 screen equivalent to 5 in screen at 17 in, Audio speaker                               & WiFi 802.11 dual-band a/b/g/n/ac, Bluetooth 4.1/2.1, microUSB          \\ 
		Vuzix M100                          & Camera (5MP still/1080p video), proximity (inward and outward), gyroscope, accelerometer, compass, 3-DoF head tracking, light, dual microphones & 4 buttons, remote control via app (iOS, Android), voice, gestures                                                                   & TI OMAP 4460 1.2GHz                                                  & Android 4.04                           & 16:9 WQVGA (equivalent to 4 in at 14 in), Audio speaker                                     & WiFi 802.11 b/g/n, Bluetooth, microUSB                                 \\ 
		ODG R-7                             & Camera (1080p @ 60fps, 720p @ 120fps), 2 accelerometer, 2 gyroscope, 2 magnetometer, altitude, humidity, light, 2 microphones                   & at least 2 buttons, trackpad, Bluetooth keyboard                                                                                    & Qualcomm SnapdragonTM 805 2.7GHz quad-core                          & ReticleOS (Android-based) & Dual 720p 16:9 stereoscopic see-thru displays, Stereo Audio ports with earbuds              & WiFi 802.11ac, Bluetooth 4.1, GNSS (GPS/GLONASS), microUSB             \\ 
		Epson Moverio BT-300                & Camera (5MP with motion tracking), magnetic, acceleromter, gyroscope, microphone                                                                & trackpad (tethered)                                                                                                                 & Intel Atom 5, 1.44GHz quad-core                                      & Android 5.1                                       & Binocular 1280x720 transparent display, Earphones                                           & WiFi 802.11 dual-band a/b/g/n/ac, Bluetooth, GPS, microUSB             \\ 
		HTC Vive*                           & Camera, gyroscope, accelerometer, laser position sensor, microphones                                                                            & Two handheld controllers with trackpad, triggers; two external base stations for tracking & (directly via host computer, GPU recommended) & Windows for host computer (SteamVR 3D engine)     & Binocular 1080x1200 (110$\degree$ FoV), Audio jack                                          & HDMI or Display Port and USB 3.0                                       \\ 
		Meta 2 Glasses                      & Camera (720p), hand tracking, 3 microphones                                                                                                     & hand gestures                                                                                                                       & (directly via host computer, GPU recommended) & Windows for host computer (Unity SDK 3D engine)   & 2550x1440 (90$\degree$ FoV), 4 surround sound speakers                                      & HDMI                                                                   \\ 
		Microsoft HoloLens                  & 4 Environment understanding cameras, depth camera, 2MP camera, 1 MR capture camera, inertial measurement unit, light, 4 microphones           & buttons, Voice and gestures                                                                                                         & Intel 32-bit, and a dedicated holographic processing unit (HPU)      & Windows 10                                        & 16:9 see-through holographic lenses (30$\degree$ FoV), Audio (surround) speakers, audio jack & WiFi 802.11ac, Bluetooth 4.1 LE, microUSB                              \\ \bottomrule
	\end{tabular}
\end{table}

The other type is the \textit{optical see-through} HMD which uses specialised optics to display objects within the user's FOV. The special optics are configured in such a way to allow light to pass through and let the users see their physical surroundings unlike the displays on the video see-through types. Epson's Moverio BT-300\footnote{Moverio BT-300 Smart Glasses (AR/Developer Edition), (Accessed 4 Aug 2017), https://epson.com/For-Work/Wearables/Smart-Glasses/Moverio-BT-300-Smart-Glasses-\%28AR-Developer-Edition\%29-/p/V11H756020} provides binocular visual augmentations using a prism projection mechanism and accepts user inputs using a capacitive muti-touch pad tethered to the eye-wear. ODG's R7\footnote{R-7 Smartglasses System, (Accessed 4 Aug 2017), https://shop.osterhoutgroup.com/products/r-7-glasses-system} smart glasses is another optical see-through device but accepts user inputs using conventional approaches, i.e. mouse and keyboard. In terms of natural user gesture support, the leading devices are the Meta 2 glasses and Microsoft's Hololens. The Meta 2\footnote{The Meta 2: Made for AR App Development, (Accessed 4 Aug 2017), https://buy.metavision.com/} delivers 3D holographic objects and utilizes environment tracking to allow a virtual 360\degree interface, and allows users to drag, pull, and drop objects within the space. However, it is tethered to a dedicated machine which has to have a dedicated graphics processor. Microsoft's HoloLens provides a similar experience to that of the Meta 2 but untethered. In early 2016, Microsoft started shipping out developer editions, and they expect that developers will create numerous services and applications around the HoloLens \cite{hololens}. The Hololens\footnote{HoloLens hardware details, (Accessed 4 Aug 2017), https://developer.microsoft.com/en-us/windows/mixed-reality/hololens\_hardware\_details/}
packs two high-definiton light engines, holographic lenses, an inertial measurement unit (IMU), 4 environment understanding cameras, a depth camera, and a dedicated holographics processing unit aside from the regular hardware (e.g. processor, memory, and battery) that the other devices have.

\section{Mixed Reality Software}\label{asec:software}

\begin{table}
	\centering
	\scriptsize
	\caption{Software Platforms, Toolkits, and Application Programming Interfaces}
	\label{tab:toolkits}
	\setlength{\extrarowheight}{.5em}
	\begin{tabular}{p{2cm}p{2.5cm}p{2.5cm}p{1.5cm}p{1.5cm}p{1.8cm}}
		\toprule
		\textbf{Software Tool Kit} &\textbf{Detection or Tracking}                                      &\textbf{Platform}            &\textbf{Development Language} & \textbf{Licensing} &\textbf{Current Version} \\ \hline
		ARToolKit                                                                                & 2D objects                                                              & Android, iOS, Linux, macOS, Unity3D                              & C, C++, C\#, Java                                  & Free (open source)                      & ARToolKit6 (2017)                             \\ 
		Layar                                                                                    & 2D objects                                           & Android, iOS, Blackberry                                         & HTTP (RESTful), JSON             & Paid                                    & Layar v7.0                                    \\ 
		Vuforia                                                                                  & 3D objects, marker-based visual SLAM                                          & Android, iOS, Windows (selected devices), Unity 3D               & C++, C\#, Objective-C, Java                        & Free (with paid versions)               & Vuforia 6.5                                   \\ 
		Zapbox                                                                                   & 3D objects, marker-based visual SLAM                                   & Android, iOS                                                     & Javascript                                         & Paid (with 30-day free trial)           & Zapworks Studio (2017)                        \\ 
		EasyAR                                                                                   & 3D objects, visual SLAM                                            & Android, iOS, macOS, Windows, Unity3D                            & C, C++, Objective-C, Java                          & Free (with paid version)                 & EasyAR SDK 2.0                                \\ 
		kudan                                                                                    & 3D object, visual SLAM                                    & Android, iOS, Unity 3D                                           & Objective-C, Java                                  & Free (with paid versions)               & kudan v1.0                                    \\ 
		Wikitude                                                                                 & 3D objects, visual SLAM                                    & Android, iOS, Smart Glasses (Epson Moverio, ODG, Vuzix), Unity3D & Objective-C, Java, Javascript                      & Paid                                    & wikitude SDK 7.0                              \\ 
		ARkit                                                                                    & 3D objects, visual and depth SLAM               & iOS                                                              & Objective-C                                        & Free                                    & ARKit (2017)                                  \\ 
		OSVR                                                                                     & 2D objects, orientation-based SLAM  & Vuzix, OSVR, HTC Vive, Oculus, SteamVR                           & C, C++, C\#, JSON for plug-ins)                   & Free (open source)                      & OSVR Hacker Development Kit (HDK) 2.0         \\ 
		Meta 2 SDKs                                                                              & 3D objects, visual and depth SLAM    & Meta 2 glasses                                                   & C\#, C++, Javascript                               & Free                                    & Meta 2 SDKs 
		\\ 
		Windows Mixed Reality                                                                    & 3D objects; visual, depth, and orientation-based SLAM   & Windows MR and HoloLens, Unity                                   & C\#, C++, Javascript              & Free                                    & Windows MR API                                \\ \bottomrule
	\end{tabular}
\end{table}

Table \ref{tab:toolkits} lists some popular software tool kits and platforms. Most of these are multi-operating system and vision-based. Specifically, they are utilizing monocular (single vision) detection for 2D, 3D , and (for some) synchronous localization and mapping (or SLAM). Of these multi-OS, vision-based tool kits, only the ARToolkit\footnote{ARToolKit6, (Accessed August 4, 2017), https://artoolkit.org/} is free and open source. Most of the paid tool kits, on the other hand, have cloud support and are primarily focused on content and service creation, and application development. Furthermore, most of these tool kits such as Wikitude\footnote{wikitude, (Accessed August 4, 2017), https://www.wikitude.com/}, Layar\footnote{Layar, (Accessed August 4, 2017), https://www.layar.com/}, and Zapbox\footnote{Zapbox: Mixed Reality for Everyone, (Accessed August 4, 2017), https://www.zappar.com/zapbox/} are also shipped as platform applications that can be used to run other applications developed using the same tool kit.

There are also platform- and device-dependent tool kits listed. The ARkit\footnote{ARKit - Apple Developer, (Accessed August 4, 2017), https://developer.apple.com/arkit/} is specifically for Apple's devices. The OSVR (Open Source Virtual Reality)\footnote{What is OSVR?, (Accessed August 4, 2017), http://www.osvr.org/what-is-osvr.html} is primarily for VR applications of specific smart glass devices, e.g HTC Vive and some Vuzix glasses, but can also be used for AR functionalities if devices have environment sensing capabilities such as the front-facing cameras of the Vive. The Meta 2 SDKs\footnote{Meta 2 SDK Features, (Accessed August 4, 2017), https://www.metavision.com/develop/sdkfeatures} are obviously for the Meta 2 glasses. Similarly, the Windows MR APIs\footnote{Windows Mixed Reality, (Accessed August 4, 2017), https://developer.microsoft.com/en-us/windows/mixed-reality} are for the Windows MR devices and the Hololens. These tool kits are primarily designed for the functionalities of these devices which includes other environmental sensing capabilities, e.g. depth sensing and head tracking, aside from the regular camera-based sensing.

\section{Summarizing Security and Privacy Strategies}\label{asec:strategies}

\begin{table}
	\centering
	\caption{Summary of MR approaches that have been discussed, and which security and privacy controls are applied by each approach and to what extent.}
	\label{tab:approaches}
	\tiny
	\vspace{-2mm}
	\begin{tabular}{@{}p{4.65cm}p{4.3mm}p{4.3mm}p{4.3mm}p{4.3mm}p{4.3mm}p{4.3mm}p{4.3mm}p{4.3mm}p{4.3mm}p{4.3mm}p{4.3mm}p{4.3mm}p{6mm}@{}}
		\toprule
		\textbf{Approach}   &    \rotatebox{60}{\parbox{1.8cm}{Resource\newline Access Control}} &   \rotatebox{60}{\parbox{1.8cm}{Device Identification\newline \& Authenticaiton}}    & \rotatebox{60}{\parbox{1.8cm}{Media Protection\newline Policy}}    & \rotatebox{60}{\parbox{1.8cm}{Media\newline Access Control}}    & \rotatebox{60}{\parbox{1.8cm}{Media Sanitization}}     & \rotatebox{60}{\parbox{1.8cm}{Physical\newline Access Control}}    & \rotatebox{60}{\parbox{1.8cm}{Information Sharing,\newline Shared Resources}}    & \rotatebox{60}{\parbox{1.8cm}{Transmission\newline Confidentiality}}    & \rotatebox{60}{\parbox{1.8cm}{Cryptographic}}    & \rotatebox{60}{\parbox{1.8cm}{Distributed\newline Processing}}    & \rotatebox{60}{\parbox{1.8cm}{Out-of-band\newline Channel}} & \rotatebox{60}{\parbox{1.8cm}{Probabilistic}}  \\ \midrule
		\multicolumn{11}{c}{{ Input Protection Approaches}} \\ \midrule
		\textsc{Darkly }\cite{jana2013scanner}$\ddagger$   &  $\square$      & $\square$      & $\blacksquare$ & $\boxplus$*        & $\blacksquare$ & $\square$      & $\square$      & $\square$      & $\square$      & $\square$      & $\square$      & $\square$ \\
		\textit{Context-based sanitization }\cite{zarepour2016context}$\ddagger$  & $\square$     & $\square$      & $\boxplus$    & $\square$      & $\blacksquare$ & $\square$      & $\square$      & $\square$      & $\square$      & $\square$      & $\square$      & $\square$ \\
		\textsc{PlaceAvoider }\cite{templeman2014placeavoider}$\dagger$  &  $\square$    & $\square$      & $\boxplus$    & $\square$      & $\blacksquare$ & $\square$      & $\square$      & $\square$      & $\square$      & $\square$      & $\square$      & $\square$ \\
		\textit{3D humanoids replace humans }\cite{szczuko2014augmented}$^{\circ}$  & $\square$        & $\square$      & $\boxplus$    & $\square$       & $\blacksquare$ & $\square$     & $\square$      & $\square$      & $\square$      & $\square$      & $\square$        & $\square$ \\
		\textsc{OpenFace/RTFace }\cite{wang2017scalable}$\dagger$   & $\square$       & $\square$      & $\boxplus$    & $\square$       & $\blacksquare$ & $\square$      & $\square$      & $\square$      & $\square$      & $\square$      & $\square$      & $\square$ \\
		\textit{Capture-resisitant spaces }\cite{truong2005preventing}$^{\circ}$   & $\square$    & $\square$      & $\square$      & $\square$      & $\square$     & $\blacksquare$   & $\square$     & $\square$      & $\square$      & $\square$      & $\square$           & $\square$ \\
		\textit{See-through vision }\cite{hayashi2010installation}$\ddagger$  &  $\square$  & $\square$      & $\square$      & $\square$    & $\blacksquare$     & $\square$     & $\square$      & $\square$      & $\square$      & $\square$      & $\square$      & $\square$ \\
		\textit{World-driven }\cite{roesner2014world}$\ddagger$  & $\square$ & $\square$      & $\boxplus$    & $\boxplus$     & $\blacksquare$ & $\square$      & $\square$      & $\square$      & $\blacksquare$ & $\square$      & $\square$      & $\square$ \\
		\textsc{I-pic }\cite{aditya2016pic}$\dagger$   &  $\square$    & $\square$      & $\boxplus$    & $\boxplus$     & $\blacksquare$ & $\square$      & $\square$      & $\square$      & $\square$      & $\square$      & $\square$      & $\square$ \\
		\textsc{PrivacyCamera }\cite{li2016privacycamera}$\dagger$   &  $\square$  & $\square$      & $\boxplus$    & $\boxplus$     & $\blacksquare$ & $\square$      & $\square$      & $\square$      & $\square$      & $\square$      & $\square$      & $\square$ \\
		\textsc{Cardea }\cite{shu2016cardea}$\dagger$    & $\square$  & $\square$      & $\boxplus$    & $\boxplus$    & $\blacksquare$ & $\square$      & $\square$      & $\square$      & $\square$      & $\square$      & $\square$      & $\square$ \\
		\textsc{MarkIt }\cite{raval2014markit,raval2016you}$\dagger$  &  $\square$  & $\square$      & $\boxplus$    & $\boxplus$     & $\blacksquare$ & $\square$      & $\square$      & $\square$      & $\square$      & $\square$      & $\square$      & $\square$ \\
		\textsc{PrePose }\cite{figueiredo2016prepose}$\ddagger$   &  $\blacksquare$*    & $\square$      & $\square$      & $\square$     & $\square$      & $\square$      & $\square$      & $\square$      & $\square$      & $\square$      & $\square$      & $\square$ \\
		\textsc{Recognizers }\cite{jana2013enabling}$\ddagger$  &  $\blacksquare$*   & $\square$      & $\blacksquare$ & $\boxplus$     & $\blacksquare$ & $\square$      & $\square$      & $\square$      & $\square$      & $\square$      & $\square$      & $\square$ \\ \midrule
		\multicolumn{11}{c}{{ Data Protection Approach/es}} \\ \midrule		
		\textsc{Rappor }\cite{erlingsson2014rappor}$^{\circ}$        & $\square$  & $\square$     & $\square$      & $\square$      & $\square$      & $\square$      & $\square$      & $\square$      & $\square$      & $\boxplus$      & $\square$      & $\blacksquare$ \\
		\textsc{SemaDroid }\cite{xu2015semadroid}$^{\circ}$        & $\blacksquare$  & $\square$     & $\square$      & $\square$      & $\square$      & $\square$      & $\square$      & $\square$      & $\square$      & $\square$      & $\square$      & $\square$ \\
		\textsc{HE-sift }\cite{jiang2017secure}$^{\circ}$   &    $\square$       &     $\square$      &  $\square$        &   $\square$        &    $\square$   &   $\square$    &  $\square$        &   $\square$      &    $\blacksquare$       &    $\square$      &       $\square$        & $\square$ \\
		\textit{Leveled} \textsc{HE-sift}\cite{zhang2014cloud}$^{\circ}$    &    $\square$       &     $\square$      &  $\square$        &   $\square$        &    $\square$   &   $\square$    &  $\square$        &   $\square$      &    $\boxplus$       &    $\square$      &       $\square$        & $\square$ \\
		\textsc{CryptoImg }\cite{ziad2016cryptoimg}$^{\circ}$      &    $\square$       &     $\square$      &  $\square$        &   $\square$        &    $\square$   &   $\square$    &  $\square$        &   $\square$      &    $\blacksquare$       &    $\square$      &       $\square$        & $\square$ \\
		\textsc{SecSift }\cite{qin2014private,qin2014towards,qin2014privacy}$^{\circ}$      &    $\square$       &     $\square$      &  $\square$        &   $\square$        &    $\square$   &   $\square$    &  $\square$        &   $\square$      &    $\blacksquare$       &    $\square$      &       $\square$        & $\square$ \\
		\textsc{P3 }\cite{ra2013p3}$^{\circ}$    &    $\square$       &     $\square$      &  $\square$        &   $\square$        &    $\square$   &   $\square$    &  $\square$        &   $\square$      &    $\boxplus$       &    $\square$      &       $\square$        & $\square$ \\
		\textit{Cloth try-on }\cite{sekhavat2017privacy}$\ddagger$  &  $\square$  & $\square$      & $\boxplus$    & $\square$      & $\square$      & $\square$      & $\square$      & $\boxplus$      & $\blacksquare$ & $\blacksquare$ & $\square$      & $\square$ \\
		\textsc{PDV }\cite{mun2010pdv}$^{\circ}$                                 & $\square$             & $\square$      & $\boxplus$    & $\boxplus$   & $\square$      & $\square$      & $\square$      & $\boxplus$    & $\square$   & $\square$  & $\square$      & $\square$ \\
		\textsc{OpenPDS }\cite{deMontjoye2014openpds}$^{\circ}$                      & $\square$       & $\square$      & $\boxplus$    & $\blacksquare$ & $\square$      & $\square$      & $\square$      & $\blacksquare$ & $\square$    & $\square$  & $\square$     & $\square$ \\
		\textsc{DataBox }\cite{crabtree2016databox}$^{\circ}$                        & $\square$      & $\square$      & $\boxplus$    & $\blacksquare$ & $\square$      & $\square$      & $\square$      & $\blacksquare$ & $\square$   & $\square$  & $\square$      & $\square$ \\ \midrule
		\multicolumn{11}{c}{Output Protection Approaches} \\ \midrule
		\textsc{Arya }\cite{lebeck2016safely, lebeck2017securing}$\ddagger$    &  $\blacksquare$ & $\square$       & $\square$      & $\square$        & $\square$      & $\square$      & $\square$      & $\square$      & $\square$      & $\square$      & $\square$      & $\square$ \\
		\textsc{SurroundWeb }\cite{vilk2014least, vilk2015surroundweb}$\ddagger$   & $\boxplus$*  & $\square$       & $\square$      & $\square$    & $\square$      & $\square$      & $\square$      & $\square$      & $\square$      & $\square$      & $\square$      & $\square$ \\  
		\textsc{EyeGuide }\cite{eaddy2004my}$\ddagger$   &  $\square$    & $\square$      & $\square$      & $\square$   & $\square$      & $\blacksquare$ & $\boxplus$    & $\square$      & $\square$      & $\square$      & $\square$      & $\square$ \\
		\textsc{VR Codes }\cite{woo2012vrcodes}$\ddagger$  & $\square$   & $\square$      & $\square$      & $\square$    & $\square$      & $\boxplus$    & $\boxplus$    & $\square$      & $\square$      & $\square$      & $\square$      & $\square$ \\
		\textit{Psycho-visual modulation }\cite{lin2017video}$\ddagger$  & $\square$   & $\square$      & $\square$      & $\square$     & $\square$      & $\boxplus$    & $\boxplus$    & $\square$      & $\square$      & $\square$      & $\square$      & $\square$ \\
		\textit{Visual cryptography using AR }\cite{lantz2015visual}$\ddagger$   &  $\square$    & $\square$      & $\square$      & $\square$   & $\square$      & $\square$      & $\blacksquare$ & $\blacksquare$ & $\boxplus$      & $\square$      & $\square$      & $\square$ \\
		\textit{Secret messages using AR }\cite{andrabi2015usability}$\ddagger$  &$\square$     & $\square$      & $\square$      & $\square$   & $\square$      & $\square$      & $\blacksquare$ & $\blacksquare$ & $\boxplus$      & $\square$      & $\square$      & $\square$ \\
		\textit{Mobile visual cryptography }\cite{fang2010securing}$\ddagger$  & $\square$     & $\square$      & $\square$      & $\square$    & $\square$      & $\square$      & $\blacksquare$ & $\blacksquare$ & $\boxplus$      & $\square$      & $\square$      & $\square$ \\
		\midrule
		\multicolumn{11}{c}{Interaction Protection Approaches}   \\ \midrule
		\textsc{Emmie }\cite{butz1998vampire, butz1999emmie}$\ddagger$  & $\square$      & $\square$      & $\square$      & $\square$      & $\square$      & $\square$      & $\blacksquare$ & $\square$      & $\square$      & $\boxplus$    & $\square$      & $\square$ \\
		\textsc{Kinected Conferences }\cite{devincenzi2011kinected}$\ddagger$  & $\square$  & $\square$      & $\square$      & $\square$    & $\square$      & $\square$      & $\blacksquare$ & $\square$      & $\square$      & $\boxplus$    & $\square$      & $\square$ \\
		\textsc{SecSpace }\cite{reilly2014secspace}$\ddagger$  & $\square$ & $\square$      & $\square$      & $\square$    & $\square$      & $\square$      & $\blacksquare$ & $\square$      & $\square$      & $\boxplus$    & $\square$      & $\square$ \\
		\textit{Candid Interactions }\cite{ens2015candid}$^{\circ}$        & $\square$      & $\square$      & $\square$      & $\square$      & $\square$      & $\square$      & $\boxplus$    & $\square$      & $\square$      & $\square$      & $\boxplus$    & $\square$ \\
		\textsc{Pip }\cite{szalavari1997personal, szalavari1998gaming}$\ddagger$ &  $\square$   & $\square$      & $\square$      & $\square$        & $\square$      & $\square$      & $\boxplus$    & $\square$      & $\square$      & $\boxplus$    & $\square$      & $\square$ \\
		\textsc{TouchSpace }\cite{cheok2002touch}$\ddagger$ & $\square$    & $\square$      & $\square$      & $\square$      & $\square$      & $\square$      & $\boxplus$    & $\square$      & $\square$      & $\boxplus$    & $\square$      & $\square$ \\
		\textsc{BragFish }\cite{xu2008bragfish}$\ddagger$   & $\square$     & $\square$      & $\square$      & $\square$      & $\square$      & $\square$      & $\boxplus$    & $\square$      & $\square$      & $\boxplus$    & $\square$      & $\square$ \\
		\textit{LooksGoodTome }\cite{gaebel2016looks}$\dagger$   & $\square$     & $\square$      & $\square$      & $\square$      & $\square$      & $\square$      & $\blacksquare$ & $\boxplus$      & $\blacksquare$ & $\square$      & $\blacksquare$ & $\square$ \\ 
		\textsc{HoloPair }\cite{sluganovic2017holopair}$\dagger$   & $\square$     & $\square$      & $\square$      & $\square$      & $\square$      & $\square$      & $\blacksquare$ & $\boxplus$      & $\blacksquare$ & $\square$      & $\blacksquare$ & $\square$ \\  \midrule
		\multicolumn{11}{c}{Device Protection Approaches}  \\ \midrule
		\textit{Seamless and secure VR }\cite{george2017seamless}$\ddagger$  & $\square$   & $\boxplus$      & $\square$      & $\square$      & $\square$      & $\square$      & $\square$      & $\square$      & $\square$      & $\square$      & $\square$      & $\square$ \\
		\textit{Mid-air authentication gestures }\cite{aslan2014mid}$\dagger$  & $\square$  & $\blacksquare$ &  $\square$      & $\square$      & $\square$      & $\square$      & $\square$      & $\square$      & $\square$      & $\square$      & $\square$      & $\square$ \\
		\textit{Head and blinking gestures }\cite{rogers2015approach}$\dagger$  & $\square$   & $\blacksquare$    & $\square$      & $\square$      & $\square$      & $\square$      & $\square$      & $\square$      & $\square$      & $\square$      & $\square$      & $\square$ \\
		\textsc{HeadBanger }\cite{li2016move}$\ddagger$  & $\square$   & $\blacksquare$    & $\square$      & $\square$      & $\square$      & $\square$      & $\square$      & $\square$      & $\square$      & $\square$      & $\square$      & $\square$ \\
		\textsc{PSKA }\cite{venkatasubramanian2010pska}$^{\circ}$  & $\square$      & $\blacksquare$ & $\square$      & $\square$      & $\square$      & $\square$      & $\square$      & $\square$      & $\square$      & $\blacksquare$     & $\square$      & $\square$ \\
		\textsc{SkullConduct }\cite{schneegass2016skullconduct}$\dagger$  & $\square$      & $\blacksquare$   & $\square$      & $\square$      & $\square$      & $\square$      & $\square$      & $\square$      & $\square$      & $\square$      & $\square$      & $\square$ \\
		\textit{Facial multi-modal authentication }\cite{raja2015multi}$\dagger$  &$\square$   & $\blacksquare$ & $\square$          & $\square$      & $\square$      & $\square$      & $\square$      & $\square$      & $\square$      & $\square$      & $\square$      & $\square$ \\
		\textsc{GazeTouchPass }\cite{khamis2016gazetouchpass}$\dagger$   & $\square$     & $\blacksquare$   & $\square$      & $\square$      & $\square$      & $\square$      & $\square$      & $\square$      & $\square$      & $\square$      & $\square$      & $\square$ \\
		\textit{Polarization }\cite{kohno2016display}$\ddagger$   & $\square$   & $\square$      & $\square$         & $\square$      & $\square$    & $\blacksquare$      & $\square$      & $\square$      & $\square$   & $\square$  & $\square$         & $\square$ \\ 
		\textsc{Taps} Widget \cite{mollers2011taps}$^{\circ}$     & $\square$        & $\square$      & $\square$      & $\square$      & $\square$       & $\blacksquare$   & $\square$      & $\square$      & $\square$      & $\square$      & $\square$        & $\square$ \\
		\textit{Chameleon-like }\cite{pearson2017chameleon}$^{\circ}$   & $\square$       & $\square$      & $\square$      & $\square$      & $\square$   & $\blacksquare$       & $\square$      & $\square$      & $\square$      & $\square$      & $\square$     & $\square$ \\
		\textsc{EyeDecrypt }\cite{forte2014eyedecrypt}$\ddagger$   & $\square$    & $\square$      & $\square$      & $\square$     & $\square$ & $\blacksquare$ & $\blacksquare$ & $\blacksquare$ & $\blacksquare$ & $\boxplus$    & $\boxplus$    & $\square$ \\
		\textit{Preventing keystroke inference }\cite{maiti2017preventing}$\ddagger$  & $\square$   & $\square$      & $\square$      & $\square$    & $\square$      & $\boxplus$    & $\blacksquare$ & $\square$      & $\square$      & $\square$      & $\square$      & $\square$ \\
		\bottomrule
	\end{tabular}
	\begin{tabular}{p{0.95\textwidth}}
	The extent of each control was either $\blacksquare$ fully applied, or $\boxplus$ partially applied, while the blank $\square$ indicates not applied, not applicable, or not mentioned in the corresponding approach. The controls have been applied to either an $\ddagger$ MR context, a $\dagger$ proto-MR context, or a $^{\circ}$ non-MR context.
	\end{tabular}
\end{table}

The previous discussions on the different security and privacy approaches were arranged based on the categorization presented in \S\ref{ssec:categorization}. Now, we present an over-all comparison of these approaches based on which security and privacy controls have been used by these approaches. We also highlight the focus of the different approaches discussed and, consequently, reveal some shortcomings, gaps, or potential areas of further study or investigation.

For Table \ref{tab:approaches}, we use the US National Institute of Standard and Technology's (NIST) listed security and privacy controls \cite{nist} as basis to formulate eleven generic control families which we use to \textit{generalize} and \textit{compare} the approaches from a high-level perspective. The 11 security and privacy controls are as follows: \textit{resource access control} and policy; \textit{device authentication} and \textit{identification}; \textit{media protection policy}; \textit{media access}; \textit{media sanitization}; \textit{physical access control}, and access to display devices; information in \textit{shared resources} and \textit{information sharing}; \textit{transmission confidentiality} and \textit{integrity}; \textit{cryptographic} protection, key establishment and management, and/or PKI certificates; \textit{distributed processing} and \textit{storage}; and \textit{out-of-band channels}. We add an additional control family that utilises \textit{probabilistic} strategies which makes the total number of families to twelve. We manually identify the controls used by the approaches and to what extent it has been applied (indicated by the shade on the boxes).

Most of the input protection approaches are fully implementing media sanitization as a form of control, as well as media protection and access policies, to some extent. Two of them were using resource access control policies to limit or provide abstraction on how applications have access to the input resources, i.e. sensing interfaces, and they further implemented \textit{least privilege} access control.

Most of the data protection approaches discussed are non-MR, and only one of them (including the approaches in other categories) utilises probabilistic controls to provide protection. In addition, only one work as well has actually been applied in an MR context. Therefore, there is big opportunity to look into MR-targeted data protection approaches in all three major data aspects -- collection, processing, and storage -- and possibly capitalizing on probabilistic strategies.

The first three output protection approaches were partially providing protection of outputs in shared resources, and physical access control to the output interfaces. Similarly, the next three approaches were providing protection in shared resources but in a full extent, as well as transmission confidentiality and cryptographic protection. The two remaining output approaches were more focused on the access the applications have to the output resources, i.e. displays.

The protection approaches for user interactions are providing protection, primarily, in an actual sharing activity in a shared space. Thus, these interaction protection approaches are addressing both aspects of information sharing and information in shared resources. Furthermore, most interaction protection approaches are also using distributed processing controls.

Among all the approaches, only the device protection ones are using device authentication controls, while the last five device protection approaches were more focused on the physical aspect of device access.

\end{screenonly}

%
%
%
%
%
%
%
%
%
%

\bibliographystyle{ACM-Reference-Format}
\bibliography{bib_all}

\end{document}